\documentclass{article}
\usepackage[latin9]{inputenc}
\usepackage{mathtools}
\usepackage{amsmath}
\usepackage{amsthm}
\usepackage{amssymb}
\usepackage{esint}

\makeatletter
\numberwithin{equation}{section}
\theoremstyle{plain}
\newtheorem{thm}{\protect\theoremname}
  \theoremstyle{remark}
  \newtheorem*{rem*}{\protect\remarkname}
  \theoremstyle{plain}
  \newtheorem{prop}{\protect\propositionname}
  \theoremstyle{plain}
  \newtheorem{lem}{\protect\lemmaname}
  \theoremstyle{remark}
  \newtheorem{rem}{\protect\remarkname}
  \theoremstyle{definition}
  \newtheorem{defn}{\protect\definitionname}
  \theoremstyle{plain}
  \newtheorem*{thm*}{\protect\theoremname}

\usepackage{amsfonts}
\usepackage{graphicx}
\providecommand{\U}[1]{\protect\rule{.1in}{.1in}}

\makeatother

  \providecommand{\definitionname}{Definition}
  \providecommand{\lemmaname}{Lemma}
  \providecommand{\propositionname}{Proposition}
  \providecommand{\remarkname}{Remark}
  \providecommand{\theoremname}{Theorem}
\providecommand{\theoremname}{Theorem}

\begin{document}

\title{Non-commutative integrability, exact solvability \\and the Hamilton-Jacobi
theory}

\author{Sergio \ Grillo\\
{\small{}Instituto Balseiro, Universidad Nacional de Cuyo and CONICET}\\[-5pt]
{\small{}Av. Bustillo 9500, San Carlos de Bariloche}\\[-5pt] {\small{}R8402AGP,
República Argentina}}
\maketitle
\begin{abstract}
The non-commutative integrability (NCI) is a property fulfilled by
some Hamiltonian systems that ensures, among other things, the exact
solvability of their corresponding equations of motion. The latter
means that an ``explicit formula'' for the trajectories of these
systems can be constructed. Such a construction rests mainly on the
so-called \textit{Lie theorem on integrability by quadratures}. It
is worth mentioning that, in the context of Hamiltonian systems, the
NCI has been for around 40 years, essentially, the unique criterium
for exact solvability expressed in the terms of first integrals (containing
the usual Liouville-Arnold integrability criterium as a particular
case). Concretely, a Hamiltonian system with $n$ degrees of freedom
is said to be non-commutative integrable if a set of independent first
integrals $F_{1},...,F_{l}$ are known such that: the kernel of the
$l\times l$ matrix with coefficients $\left\{ F_{i},F_{j}\right\} $,
where $\left\{ \cdot,\cdot\right\} $ denotes the canonical Poisson
bracket, has dimension $2n-l$ (\textbf{isotropy}); and each bracket
$\left\{ F_{i},F_{j}\right\} $ is functionally dependent on $F_{1},...,F_{l}$
(\textbf{closure}). In this paper, we develop two procedures for constructing
the trajectories of a Hamiltonian system which only require isotropic
first integrals (closure condition is not needed). One of them is
based on an extended version of the geometric Hamilton-Jacobi theory,
and does not rely on the above mentioned Lie's theorem. We do all
that in the language of functions of several variables. 
\end{abstract}

\section{Introduction}

Consider an ordinary differential equation (ODE) of the form
\[
\dot{y}\left(t\right)=f\left(t,y\left(t\right)\right).
\]
It is well-known that the continuity of $f$ on an open subset $U\subseteq\mathbb{R}^{2}$
is enough to ensure the existence of a solution passing through each
point of $U$ (see for instance \cite{ar2,hur}). But, which is the
expression of such a solution? Can we find an explicit formula for
it? In general, we do not know how to do that. In the particular case
in which the continuous function $f$ is of the form
\[
f\left(t,y\right)=g\left(t\right)/h\left(y\right)
\]
(and consequently $h\left(y\right)\neq0$ for all $y$), the unique
solution passing through $\left(t_{0},y_{0}\right)\in U$ is given
by 
\[
\int_{y_{0}}^{y\left(t\right)}h\left(s\right)ds=\int_{t_{0}}^{t}g\left(s\right)ds,
\]
or equivalently, in terms of a primitive or \textit{quadrature} $H$
(resp. $G$) of $h$ (resp. $g$), the solution is the curve satisfying
\[
H\left(y\left(t\right)\right)=H\left(y_{0}\right)+G\left(t\right)-G\left(t_{0}\right).
\]
Since, $H'\left(y\right)=h\left(y\right)\neq0$ for all $y$, it is
clear that $H$ is injective, so above equation can be univocally
solved for $y\left(t\right)$. In any case, for the given function
$f$, we could transform the original ODE into an algebraic equation
(with the same solutions). In general, this is the best we can do
in order to solve an ODE (or a system of them). When this happens,
one uses to say that the given ODE is \textbf{exactly solvable}. And
when the data of the algebraic equation, as in the present case, is
given by primitives of the original data, one says that the solutions
can be \textbf{constructed, or integrated, up to quadratures}. Nevertheless,
we shall use the phrases ``exactly solvable'' and ``integrable
up to quadratures'' as synonyms.

\bigskip{}

In the context of Hamiltonian systems, the exact solvability of their
equations of motion is ensured by the \textit{non-commutative integrability}
property. This can be shown by using the \textit{Lie theorem on integrability
by quadratures} \cite{akn,lie} (for recent extensions of the theorem,
see \cite{pep1,pep2}). A Hamiltonian system defined by $H:\mathbb{R}^{2n}\rightarrow\mathbb{R}$
is \textbf{non-commutative integrable (NCI)}, \textbf{superintegrable}
or \textbf{Mischenko-Fomenko integrable} \cite{mf} (see also \cite{j}
and references therein), if functions $F_{1},...,F_{l}:\mathbb{R}^{2n}\rightarrow\mathbb{R}$
such that:
\begin{enumerate}
\item (\textbf{independence}) the rank of the Jacobian matrix of $F=\left(F_{1},...,F_{l}\right)$
is $l$; 
\item (\textbf{first integrals}) $\left\{ F_{i},H\right\} =0$ for all $i$;
\item (\textbf{isotropy}) the matrix with coefficients $\left\{ F_{i},F_{j}\right\} $
has a kernel of dimension $2n-l$;
\item (\textbf{closure}) for each $i,j$ there exists a function $P_{ij}:\mathsf{Im}F\subseteq\mathbb{R}^{l}\rightarrow\mathbb{R}$
such that $\left\{ F_{i},F_{j}\right\} =P_{ij}\circ\left(F_{1},...,F_{l}\right)$;
\end{enumerate}
\textit{are known}.\footnote{What is important here is not the existence of the functions $F_{1},...,F_{l}$,
but the fact that we know them. In order to emphasize that, sometimes
a NCI system is defined as a pair $\left(H,F\right)$, with $F=\left(F_{1},...,F_{l}\right)$
satisfying conditions above.} Here, $\left\{ \cdot,\cdot\right\} $ denotes the canonical Poisson
bracket. We are omitting another conditions that sometimes appear
in the definition of a NCI system, as the compactness and connectedness
of the common level sets of the functions $F_{1},...,F_{l}$. Such
additional conditions ensure certain qualitative behavior of the system,
and also some geometric properties, in which we are not interested
(see \cite{j} for a review). 

Note that the isotropy condition implies that $l\geq n$. In the particular
case in which $l=n$, we have the usual notion of integrability: \textbf{Liouville-Arnold}
or \textbf{commutative integrability (CI)} \cite{ar,lio}. In such
a case, the last two conditions collapse into the isotropy condition
only, which says that $\left\{ F_{i},F_{j}\right\} =0$ for all $i,j$.

If the functions $F_{1},...,F_{l}$ are just defined (or satisfy above
conditions) on an open subset $U\subseteq\mathbb{R}^{2n}$, we shall
say that the system is \textbf{NCI along }$U$ (and \textbf{CI along
}$U$ when $l=n$). And, if for each point of $\mathbb{R}^{2n}$ we
know an open neighborhood $U$ such that the system is NCI along $U$,
we shall say that the system is \textbf{locally NCI }(and \textbf{locally
CI} when $l=n$).

It is worth mentioning that, around every point of $\mathbb{R}^{2n}$,
excluding the critical points of the system, there always exist functions
$F_{i}$'s satisfying the conditions (1) to (4) (see the Appendix).
However, in the definition of NCI, we are not asking the existence
of such functions, but the knowledge of them. In fact, it is this
knowledge (and not just the existence) what enable us to construct
up to quadratures the trajectories of the system (\textit{via} the
above mentioned Lie's theorem).

\bigskip{}

At this point, we can formulate the following theoretical question:
are all the properties defining a NCI system essential in order to
ensure the exact solvability of a Hamiltonian system? In this work,
we show that conditions (1), (2) and (3) are enough for such a purpose.
We do that by following two different ways.
\begin{itemize}
\item Firstly we prove that, from an independent set of isotropic first
integrals, i.e. functions satisfying (1), (2) and (3), we can construct,
around \textit{almost} every point of the phase space, a set of functions
also satisfying (4). Here, by ``almost'' we mean that the construction
works on an open dense subset of $\mathbb{R}^{2n}$. This implies
that, in such a subset, the system is locally NCI and, consequently,
using the Lie's theorem, is also exactly solvable there. It is worth
mentioning that, in order to construct the trajectories of the system,
we need to use all the first integrals, not just the original isotropic
ones. 
\item Secondly, we use a generalized version of the Hamilton-Jacobi theory
\cite{gp}. In this case, we develop an alternative procedure for
constructing the trajectories of the system that only uses the given
isotropic first integrals (no need for constructing additional first
integrals). Such a procedure is an extension of the usual construction
of canonical transformations \textit{via} the Hamilton's characteristic
functions (which is the main aim of the standard Hamilton-Jacobi theory
\cite{ar,gold}). In particular, the Lie's theorem is not needed this
time. Also, the integration is ensured in the whole of the phase space
(not only along a dense subset).
\end{itemize}
Part of the content of this paper already appeared in \cite{gp},
but in that reference such a content was presented in the language
of the symplectic geometry. Here, we make a rather different presentation
(of the results and their proofs), using (when possible) only elementary
concepts of the calculus of several variables. Our aim is to reach
a more general public, with no background in differential geometry.
We can do that simply because the exact solvability is, essentially,
a local aspect of a dynamical system.

\bigskip{}

Summarizing, the main goal of the paper is two-fold:
\begin{description}
\item [{a.}] To show the following theoretical result: ``for a given Hamiltonian
system, the knowledge of a set of isotropic first integrals is enough
to integrate its equations of motion up to quadratures.'' 
\item [{b.}] To state and prove above result by using the simple language
of functions of several variables.
\end{description}
\bigskip{}

The paper is organized as follows. In Section 2 we prove the well-known
fact that local NCI implies exact solvability. Although the proof
is rather standard, we do it with some detail because we want to highlight
the kind of procedures which are involved in the construction of the
trajectories. Then, at the end of the section, we give our first proof
of the result described in the point $\mathbf{a}$ above. In Section
3 we make a brief review of the standard Hamilton-Jacobi theory, emphasizing
its relationship with the local CI. In Section 4 we present the extension
of the Hamilton-Jacobi theory that appears in Ref. \cite{gp}. Instead
of working in the context of the symplectic geometry, as in the mentioned
paper, we shall work in the simpler framework of functions of several
variables. In Section 5 we show the relationship between the extended
theory and the NCI. Finally, in Section 6, we elaborate a new procedure
for constructing (up to quadratures) the trajectories of a Hamiltonian
system, based on the above mentioned extension of the Hamilton-Jacobi
theory. This constitutes a second proof of our main result (see point
$\mathbf{a}$ again).

\bigskip{}

We shall assume that the reader is familiar with the basic ideas related
to Hamiltonian systems \cite{ar,gold} and to the calculus of several
variables. Nevertheless, below, we introduce some notation and recall
some useful concepts and results associated to those subjects. 

\bigskip{}

\textbf{Notation, conventions and some basic concepts. }
\begin{itemize}
\item Throughout this paper, all the functions will be of class $C^{\infty}$
on an open subset $A$ of some $\mathbb{R}^{m}$. For instance, if
we say that a function is left (resp. right) invertible, we shall
be assuming that it has a left (resp. right) inverse of class $C^{\infty}$.
So, if we have a left and right invertible function, then such a function
is a \textbf{diffeomorphism}: class $C^{\infty}$, bijective and with
inverse of class $C^{\infty}$.
\item Given $F:A\subseteq\mathbb{R}^{m}\rightarrow\mathbb{R}^{k}$, we shall
denote by $DF\left(\mathbf{x}\right)\in\mathsf{Mat}\left(k\times m,\mathbb{R}\right)$
the Jacobian matrix or differential of $F$ at the point $\mathbf{x}\in A$,
i.e. the $k\times m$ real matrix with coefficients
\[
\left[DF\left(\mathbf{x}\right)\right]_{ij}=\frac{\partial F_{i}}{\partial x_{j}}\left(\mathbf{x}\right),\;\;\;i=1,...,k,\;\;\;j=1,...,m,
\]
where each $F_{i}$ (resp. $x_{j}$) is a component of $F$ (resp.
$\mathbf{x}$). We shall also see $DF\left(\mathbf{x}\right)$ as
a linear transformation from $\mathbb{R}^{m}$ to $\mathbb{R}^{k}$.
If $k=1$ (i.e. $F=F_{1}$), $DF\left(\mathbf{x}\right)$ is a row
vector that we shall sometimes denote
\[
DF\left(\mathbf{x}\right)=\frac{\partial F}{\partial\mathbf{x}}\left(\mathbf{x}\right).
\]
\item By the \textbf{rank} of $F$ at $\mathbf{x}$ we shall mean the number
$\mathsf{rank}F\left(\mathbf{x}\right)\coloneqq\mathsf{dim}\left[\mathsf{Im}\left[DF\left(\mathbf{x}\right)\right]\right]$.
A set of $k$ functions $F_{1},...,F_{k}:A\subseteq\mathbb{R}^{m}\rightarrow\mathbb{R}$
is \textbf{independent} if the rank of $F\coloneqq\left(F_{1},...,F_{k}\right):A\subseteq\mathbb{R}^{m}\rightarrow\mathbb{R}^{k}$
is $k$ for all $\mathbf{x}\in A$. In other words, the linear transformation
$DF\left(\mathbf{x}\right):\mathbb{R}^{m}\rightarrow\mathbb{R}^{k}$
is surjective for all $\mathbf{x}$. It can be shown that, in such
a case, $\mathsf{Im}F\subseteq\mathbb{R}^{k}$ is an open subset.
Given a set of independent functions $F_{1},...,F_{k}$, for every
$\lambda\in\mathsf{Im}F$ we shall say that each pre-image $F^{-1}\left(\lambda\right)$
is a \textbf{manifold of dimension} $m-k$. On the other hand, given
another function $G:A\rightarrow\mathbb{R}$, we shall say that $G$
is \textbf{dependent on} $F_{1},...,F_{k}$ if there exists $P:\mathsf{Im}F\rightarrow\mathbb{R}$
such that $G=P\circ F$; and $G$ is \textbf{locally dependent on}
$F_{1},...,F_{k}$ if for each $\mathbf{x}\in A$ there exists an
open neighborhood $U$ of $\mathbf{x}$ and a function $P:F\left(U\right)\subseteq\mathbb{R}^{k}\rightarrow\mathbb{R}$
such that $\left.G\right|_{U}=P\circ\left.\left(F_{1},...,F_{k}\right)\right|_{U}$.
\item In this paper, we shall restrict ourself to Hamiltonian systems whose
phase space is contained in $\mathbb{R}^{2n}=\mathbb{R}^{n}\times\mathbb{R}^{n}$.
This is because we are only interested in local aspects of these systems.
So, the Hamiltonians will be functions $H:A\subseteq\mathbb{R}^{2n}\rightarrow\mathbb{R}$.
Nevertheless, for simplicity, we shall usually assume that $A=\mathbb{R}^{2n}$.
Denoting the points of $\mathbb{R}^{2n}$ by $\left(\mathbf{q},\mathbf{p}\right)=\left(q^{1},..,q^{n},p_{1},...,p_{n}\right)$,
the canonical equations for a Hamiltonian $H$ are
\begin{equation}
\dot{\mathbf{q}}\left(t\right)=\frac{\partial H}{\partial\mathbf{p}}\left(\mathbf{q}\left(t\right),\mathbf{p}\left(t\right)\right)\;\;\;\textrm{and}\;\;\;\dot{\mathbf{p}}\left(t\right)=-\frac{\partial H}{\partial\mathbf{q}}\left(\mathbf{q}\left(t\right),\mathbf{p}\left(t\right)\right).\label{ceq}
\end{equation}
\item Given two functions $F,G:A\subseteq\mathbb{R}^{2n}\rightarrow\mathbb{R}$,
its canonical Poisson bracket\textbf{ }$\left\{ F,G\right\} :A\subseteq\mathbb{R}^{2n}\rightarrow\mathbb{R}$
is given by
\begin{equation}
\left\{ F,G\right\} \left(\mathbf{x}\right)=DF\left(\mathbf{x}\right)\cdot J\cdot\left(DG\left(\mathbf{x}\right)\right)^{t},\label{pb}
\end{equation}
where
\begin{equation}
J=\left[\begin{array}{cc}
0_{n} & I_{n}\\
-I_{n} & 0_{n}
\end{array}\right]\in\mathsf{Mat}\left(2n\times2n,\mathbb{R}\right)\label{J}
\end{equation}
and $0_{n}$ (resp. $I_{n}$) denotes the $n\times n$ null (resp.
identity) matrix. Note that $J^{-1}=-J=J^{t}$. 
\item By a \textbf{vector field} on $A\subseteq\mathbb{R}^{m}$ we shall
mean a function $X:A\rightarrow\mathbb{R}^{m}$. A set of vector fields
$X_{1},...,X_{r}$ on $A$ is \textbf{linearly independent} if so
is the set of vectors $X_{1}\left(\mathbf{x}\right),...,X_{r}\left(\mathbf{x}\right)\in\mathbb{R}^{m}$
for all $\mathbf{x}\in A$. Given two vector fields $X$ and $Y$,
its \textbf{Lie bracket} $\left[X,Y\right]$ is the vector field given
by
\[
\left[X,Y\right]\left(\mathbf{x}\right)=X\left(\mathbf{x}\right)\cdot\left(DY\left(\mathbf{x}\right)\right)^{t}-Y\left(\mathbf{x}\right)\cdot\left(DX\left(\mathbf{x}\right)\right)^{t}.
\]
Every vector field $X$ on $A$ defines a dynamical system whose trajectories,
also called the \textbf{integral curves} of $X$, are the functions
$\gamma:I\subseteq\mathbb{R}\rightarrow A$ such that
\begin{equation}
\frac{d}{dt}\gamma\left(t\right)=X\left(\gamma\left(t\right)\right).\label{gem}
\end{equation}
Given a manifold $N\subseteq A$ defined by a function $F$, we shall
say that $X$ is \textbf{tangent to} $N$, or that $N$ is an\textbf{
invariant manifold} for $X$, if 
\begin{equation}
X\left(\mathbf{x}\right)\cdot\left(DF\left(\mathbf{x}\right)\right)^{t}=0,\;\;\;\forall\mathbf{x}\in N.\label{tanto}
\end{equation}
It can be shown that this is the same as saying that all the integral
curves of $X$ passing through $N$ are entirely contained in $N$
(for a proof, see Ref. \cite{boo}). 
\item Given a function $H:A\subseteq\mathbb{R}^{2n}\rightarrow\mathbb{R}$,
the vector field $X_{H}$ given by
\begin{equation}
X_{H}\left(\mathbf{x}\right)=-DH\left(\mathbf{x}\right)\cdot J=J\cdot\left(DH\left(\mathbf{x}\right)\right)^{t}\label{xf}
\end{equation}
is called \textbf{Hamiltonian vector field }associated to $H$. It
is easy to see {[}combining \eqref{ceq}, \eqref{gem} and \eqref{xf}{]}
that the integral curves of $X_{H}$ are exactly the trajectories
of $H$. Also, given functions $F$ and $G$ on $A\subseteq\mathbb{R}^{2n}$,
it can be shown that
\begin{equation}
\left[X_{F},X_{G}\right]=-X_{\left\{ F,G\right\} }.\label{corbra}
\end{equation}
\end{itemize}

\section{NCI, isotropy and Lie integrability}

In this section, in the first place, we give a proof of the well-known
fact that local NCI implies exact solvability. We do that for later
convenience, in order to highlight the kind of procedures involved
in the construction of the trajectories of a Hamiltonian system. This
will enable us to compare the different integration procedures that
appear along the paper. 

Secondly, we present the first proof of our main result: one of the
conditions appearing in the definition of NCI, the closure condition,
is no needed for ensuring exact solvability.

\subsection{From NCI to exact solvability}

\label{fnci}

Let us show that a locally NCI system is exactly solvable. The proof
will be based on the theorem below. Before stating and proving it,
let us introduce some terminology. 

We shall say that ``a function $F:A\subseteq\mathbf{\mathbb{R}}^{m}\rightarrow\mathbf{\mathbb{R}}^{k}$
\textbf{can be constructed}'' if its domain $A$ and their values
$F\left(\mathbf{x}\right)$ (for all $\mathbf{x}\in A$)
\begin{itemize}
\item are simply known;
\item they can be determined by making a finite number of arithmetic operations
(as the calculation of a determinant) and/or solving a finite set
of linear equations (which actually can be reduced to arithmetic operations);
\item or they can be expressed in terms of the derivatives and/or lateral
inverses of another (known) functions.

If the expression of $F$ also involves primitives of another functions,
we shall say that ``$F$ \textbf{can be constructed up to quadratures}.''

\end{itemize}
\begin{thm}
\textbf{(Lie integrability by quadratures) }Given a vector field $X$
on $\mathbb{R}^{m}$ tangent to an $r$-dimensional manifold $N\subseteq\mathbb{R}^{m}$,
if we know $r$ linearly independent vector fields $X_{1},...,X_{r}$
tangent to $N$ such that $\left[X_{i},X_{j}\right]\left(\mathbf{x}\right)=\left[X_{i},X\right]\left(\mathbf{x}\right)=0$
for all $i,j$ and all $\mathbf{x}\in N$, then the integral curves
of $X$ contained inside $N$ can be constructed up to quadratures.
\end{thm}
\textit{Proof}. Given $p\in N$, if $X\left(p\right)=0$, then the
integral curve through $p$ is the constant function. So, let us assume
that $X\left(p\right)\neq0$. We are going to construct (up to quadratures),
around $p$, a set of local coordinates $\left(y_{1},...,y_{r}\right)$
for the manifold $N$ where the equations of motion adopt the form\footnote{It is well-known that this kind of coordinates always exist around
non-critical points of any vector field. What is important here it
is not their existence, but the fact that they can be constructed.}
\begin{equation}
\dot{y}_{1}\left(t\right)=1,\;\dot{y}_{2}\left(t\right)=\cdots=\dot{y}_{r}\left(t\right)=0.\label{coo}
\end{equation}

Since the vector fields $X,X_{1},...,X_{r}$ are tangent to $N$,
$N$ is $r$-dimensional and the $X_{i}$'s are independent, then
we can write $X\left(p\right)$ as a linear combination of the vectors
$X_{i}\left(p\right)$'s. Since $X\left(p\right)\neq0$, such a linear
combination must have some non-null coefficient. Let us assume that
the first coefficient is non-null (otherwise, we can reorder the vector
fields). This means that the vectors $X\left(p\right),X_{2}\left(p\right),...,X_{r}\left(p\right)$
are independent. By continuity, there exists an open neighborhood
$U\subseteq N$ of $p$ where the vector fields $X,X_{2},...,X_{r}$
are independent. From now on, let us write $X=X_{1}$. 

Let $\left(x_{1},...,x_{r}\right)$ be local coordinates for $N$
defined on $U$ (shrinking $U$ if needed). Since the vector fields
$X_{1},...,X_{r}$ are tangent to $N$, the related directional derivatives
of a function $f:U\rightarrow\mathbb{R}$ can be written
\[
X_{i}\cdot\left(Df\right)^{t}=\sum_{j=1}^{r}b_{ij}\,\frac{\partial f}{\partial x_{j}},
\]
for certain functions $b_{ij}:U\subseteq N\rightarrow\mathbb{R}$.
Now, for each $k=1,...,r$, consider the equations 
\[
\sum_{j=1}^{r}b_{ij}\,\frac{\partial f}{\partial x_{j}}=\delta_{ik},\;\;\;i=1,...,r,
\]
being $\delta_{ik}$ the Kronecker delta. Since the vector fields
$X_{i}$'s are independent along $U$, then the matrix with coefficients
$b_{ij}$'s must be invertible. So, last equations are equivalent
to
\begin{equation}
\frac{\partial f}{\partial x_{i}}=\left(b^{-1}\right)_{ik},\;\;\;i=1,...,r.\label{me}
\end{equation}
On the other hand, it is easy to show that condition $\left[X_{i},X_{j}\right]=0$
is equivalent to
\[
\frac{\partial}{\partial x_{j}}\left(b^{-1}\right)_{ik}=\frac{\partial}{\partial x_{i}}\left(b^{-1}\right)_{jk},\;\;\;i,j,k=1,...,r,
\]
what implies that Equations \eqref{me} can be solved by quadratures.
In fact, for each $k$, the general solution $y_{k}$ is given by
the formula
\[
y_{k}\left(x_{1},...,x_{r}\right)=\sum_{i=1}^{r}\intop_{x_{0,i}}^{x_{i}}\left(b^{-1}\right)_{ik}\left(x_{0,1},...,x_{0,i-1},t,x_{i+1},...,x_{r}\right).
\]
We can choose the numbers $x_{0,i}$'s as the coordinates of $p$.
In such a case, it is clear that the functions $y_{1},...,y_{r}$
define a new coordinate system of $N$ around $p$. In particular,
they are independent. Moreover, they satisfy
\[
X_{1}\cdot\left(Dy_{k}\right)^{t}=\delta_{1k},\;\;\;k=1,...,r,
\]
what implies precisely Eq. \eqref{coo}. Then, the integral curves
$\left(x_{1}\left(t\right),...,x_{r}\left(t\right)\right)$ of the
field $X=X_{1}$ around $p$ are given by the algebraic equations
\[
y_{1}\left(x_{1}\left(t\right),...,x_{r}\left(t\right)\right)=t+y_{0,1},\;\;\;,y_{j}\left(x_{1}\left(t\right),...,x_{r}\left(t\right)\right)=y_{0,j},
\]
for $j=2,...,r$, which can be univocally solved for the $x_{i}\left(t\right)$'s
because the functions $y_{i}$ are independent. Since all that can
be done around any point of $N$, the theorem is proved$.\;\;\;\diamondsuit$

\bigskip{}

\begin{rem*}
As we said in the Introduction, along all of this paper, the phrases
``\textbf{the system is exactly solvable}'' and ``\textbf{the trajectories
of the system can be constructed up to quadratures}'' will be used
as synonyms.
\end{rem*}
Now, suppose that we have a NCI system (as defined in the Introduction)
with Hamiltonian function $H:\mathbb{R}^{2n}\rightarrow\mathbb{R}$
and independent first integrals $F_{1},...,F_{l}$. 

\bigskip{}

\textbf{Example 1.} As example of a (local) NCI system, we can consider
the \textit{isotropic harmonic oscillator} with $3$ degrees of freedom.
Its Hamiltonian function $H:\mathbb{R}^{6}\rightarrow\mathbb{R}$
is given by
\[
H\left(\mathbf{q},\mathbf{p}\right)=p_{1}^{2}+p_{2}^{2}+p_{3}^{2}+q_{1}^{2}+q_{2}^{2}+q_{3}^{2}.
\]
Beside $H$, the functions $H_{i},P_{ij}:\mathbb{R}^{6}\rightarrow\mathbb{R}$,
$1\leq i,j\leq3$, with 
\[
H_{i}\left(\mathbf{q},\mathbf{p}\right)=p_{i}^{2}+q_{i}^{2}\;\;\;\textrm{and}\;\;\;P_{ij}\left(\mathbf{q},\mathbf{p}\right)=q_{i}\,p_{j}-q_{j}\,p_{i},
\]
are also first integrals for the system. It can be shown that $H,H_{1},H_{2},P_{12}$
and $P_{13}$ are independent inside an open dense subset $A\subseteq\mathbb{R}^{6}$,
and for dimensional reasons they must be isotropic (see point (3))
and must satisfy the closure condition (see point (4)). So, the system
is NCI along $A$. In this case, since the number of independent first
integrals is equal to $2n-1$, it says that the system is \textit{maximally
superintegrable}$.\;\;\;\diamondsuit$

\bigskip{}

Define $F\coloneqq\left(F_{1},...,F_{l}\right)$ and for each $\lambda\in\mathsf{Im}F$
(the range of $F$) consider the level set 
\[
F^{-1}\left(\lambda\right)=\left\{ \mathbf{x}\in\mathbb{R}^{2n}:F\left(\mathbf{x}\right)=\lambda\right\} .
\]
Note that:
\begin{itemize}
\item $\mathbb{R}^{2n}$ is a (disjoint) union of the subsets $F^{-1}\left(\lambda\right)$; 
\item (independence) each subset $F^{-1}\left(\lambda\right)$ is a manifold
of dimension $r\coloneqq2n-l$;
\item (first integrals) each trajectory of the system is contained inside
some level set $F^{-1}\left(\lambda\right)$, i.e. the Hamiltonian
vector field $X_{H}$ is tangent to each manifold $F^{-1}\left(\lambda\right)$.
\end{itemize}
As a consequence, in order to find all the trajectories of our system,
it is enough to look for them on each $r$-dimensional manifold $F^{-1}\left(\lambda\right)$. 
\begin{prop}
Under above conditions and notation, for each $\lambda_{0}\in\mathsf{Im}F$,
a set of vector fields $X_{1},...,X_{r}$ on $\mathbb{R}^{2n}$ tangent
to $F^{-1}\left(\lambda_{0}\right)$ and such that
\[
\left[X_{i},X_{j}\right]\left(\mathbf{x}\right)=\left[X_{i},X_{H}\right]\left(\mathbf{x}\right)=0,\;\;\;\forall\mathbf{x}\in F^{-1}\left(\lambda_{0}\right),
\]
can be constructed.
\end{prop}
\textit{Proof}. The closure condition {[}see point (4) above{]} says
that
\[
\left\{ F_{i},F_{j}\right\} \left(\mathbf{x}\right)=P_{ij}\left(\lambda\right),\;\;\;\forall\mathbf{x}\in F^{-1}\left(\lambda\right),
\]
for some functions $P_{ij}$. Note that the last equation determines
completely each function $P_{ij}$. So, the functions $P_{ij}$'s
are known. On the other hand, the isotropy condition {[}see point
(3){]} ensures, for each $\lambda$, the existence of $r=2n-l$ linearly
independent vectors $\mathbf{v}_{1}^{\lambda},...,\mathbf{v}_{r}^{\lambda}\in\mathbb{R}^{l}$
such that
\[
\sum_{j=1}^{l}P_{ij}\left(\lambda\right)\,\left(\mathbf{v}_{k}^{\lambda}\right)_{j}=0,\;\;\;i=1,...,l,\;\;\;k=1,...,r.
\]
Moreover, it can be shown that, given $\lambda_{0}\in\mathsf{Im}F$,
we can find a neighborhood $V$ of $\lambda_{0}$ and functions (of
class $C^{\infty}$)
\begin{equation}
\lambda\in V\mapsto\left(\mathbf{v}_{k}^{\lambda}\right)_{j}\in\mathbb{R},\;\;\;j=1,...,l,\label{ul}
\end{equation}
satisfying above equation. (We just have to make standard linear manipulations).
Consider the related vector fields {[}see Eq. \eqref{xf}{]}
\begin{equation}
X_{i}^{\lambda}\coloneqq\sum_{j=1}^{l}\left(\mathbf{v}_{i}^{\lambda}\right)_{j}\,X_{F_{j}},\;\;\;i=1,...,r.\label{xli}
\end{equation}
It is easy to see that they are linearly independent. In addition,
using the point (2) and Eq. \eqref{corbra}, we have
\[
-\left[X_{F_{j}},X_{H}\right]=X_{\left\{ F_{j},H\right\} }=0,\;\;\;j=1,...,l,
\]
what implies that
\begin{equation}
\left[X_{i}^{\lambda},X_{H}\right]=0,\;\;\;i=1,...,r.\label{xh}
\end{equation}
And, since
\[
\left[X_{F_{a}},X_{F_{b}}\right]\left(\mathbf{x}\right)=-\sum_{k=1}^{l}\frac{\partial P_{ab}}{\partial\lambda_{k}}\left(F\left(\mathbf{x}\right)\right)\,X_{F_{k}}\left(\mathbf{x}\right)
\]
{[}using again \eqref{xf} and \eqref{corbra}{]}, then
\[
\left[X_{i}^{\lambda},X_{j}^{\lambda}\right]\left(\mathbf{x}\right)=-\sum_{k=1}^{l}\left(\mathbf{v}_{i}^{\lambda}\right)_{a}\,\frac{\partial P_{ab}}{\partial\lambda_{k}}\left(F\left(\mathbf{x}\right)\right)\,\left(\mathbf{v}_{j}^{\lambda}\right)_{b}\,X_{F_{k}}\left(\mathbf{x}\right).
\]
But, for all $\lambda\in U$, taking into account that $P_{ij}=-P_{ji}$,
\[
\begin{array}{lll}
\left(\mathbf{v}_{i}^{\lambda}\right)_{a}\,\frac{\partial P_{ab}}{\partial\lambda_{k}}\left(\lambda\right)\,\left(\mathbf{v}_{j}^{\lambda}\right)_{b} & = & \frac{\partial}{\partial\lambda_{k}}\left(\left(\mathbf{v}_{i}^{\lambda}\right)_{a}\,P_{ab}\left(\lambda\right)\,\left(\mathbf{v}_{j}^{\lambda}\right)_{b}\right)-\frac{\partial\left(\mathbf{v}_{i}^{\lambda}\right)_{a}}{\partial\lambda_{k}}\,P_{ab}\left(\lambda\right)\,\left(\mathbf{v}_{j}^{\lambda}\right)_{b}\\
\\
 &  & -\left(\mathbf{v}_{i}^{\lambda}\right)_{a}\,P_{ab}\left(\lambda\right)\,\frac{\partial\left(\mathbf{v}_{j}^{\lambda}\right)_{b}}{\partial\lambda_{k}}=0-0-0=0,
\end{array}
\]
so,
\begin{equation}
\left[X_{i}^{\lambda},X_{j}^{\lambda}\right]\left(\mathbf{x}\right)=0,\;\;\;\forall\mathbf{x}\in F^{-1}\left(\lambda\right).\label{xx}
\end{equation}
Finally, since
\[
X_{j}^{\lambda}\left(\mathbf{x}\right)\cdot\left(DF_{i}\left(\mathbf{x}\right)\right)^{t}=\sum_{k=1}^{l}P_{ik}\left(F\left(\mathbf{x}\right)\right)\,\left(\mathbf{v}_{j}^{\lambda}\right)_{k}=0
\]
for all $i,j$ and $\mathbf{x}\in F^{-1}\left(\lambda\right)$, then
the vector fields $X_{i}^{\lambda}$ are tangent to $F^{-1}\left(\lambda\right)$,
for all $\lambda\in V$ {[}see Eq. \eqref{tanto}{]}. Accordingly,
it is enough to define $X_{i}\coloneqq X_{i}^{\lambda_{0}}$$.\;\;\;\diamondsuit$

\bigskip{}

Combining above proposition and the Lie's theorem, it is clear that
the trajectories of a NCI system can be constructed up to quadratures.
For such a construction, we need to follow the steps below. Given
the functions $F_{1},...,F_{l}$:
\begin{enumerate}
\item construct, around each $\lambda_{0}\in\mathsf{Im}F$, the functions
given by \eqref{ul};
\item construct the vector fields $X_{1}^{\lambda_{0}},...,X_{r}^{\lambda_{0}}$
by using Eq. \eqref{xli};
\item apply the (proof of the) Lie's theorem to each manifold $F^{-1}\left(\lambda_{0}\right)$.
\end{enumerate}
If the system is just locally NCI, above construction can be made
on each open subset $U\subseteq\mathbb{R}^{2n}$ along which the system
is NCI. Since those subsets cover the whole of $\mathbb{R}^{2n}$,
again we can construct up to quadratures all the trajectories. Concluding,
\begin{thm}
\label{ncies}Every (locally) NCI system is exactly solvable.
\end{thm}

\subsection{From isotropy to NCI}

\label{isotonci}

Now, let us see that, from a set of isotropic first integrals, we
can construct another local first integrals that make the system a
locally NCI system\footnote{Perhaps the result is quite expected, but, as far as we know, its
proof is not published anywhere.} (unless on an open dense subset) and, consequently, exactly solvable.
First, we need several auxiliary results.
\begin{lem}
Consider a function $G:A\subseteq\mathbb{R}^{m}\rightarrow\mathbb{R}$
and a set of independent functions $F_{1},...,F_{k}:A\subseteq\mathbb{R}^{m}\rightarrow\mathbb{R}$.
If there exists $P:\mathsf{Im}F\subseteq\mathbb{R}^{k}\rightarrow\mathbb{R}$
such that $G=P\circ\left(F_{1},...,F_{k}\right)$, i.e. $G$ is dependent
on $F_{1},...,F_{k}$, then 
\[
\mathsf{rank}\hat{F}\left(\mathbf{x}\right)=k,\;\;\;\forall\mathbf{x}\in A,
\]
where $\hat{F}\coloneqq\left(F_{1},...,F_{k},G\right)$. Reciprocally,
if above condition holds, then the function $G$ is locally dependent
on $F_{1},...,F_{k}$. 
\end{lem}
\textit{Proof}. For the first statement, note that $k\leq\mathsf{rank}\hat{F}\left(\mathbf{x}\right)\leq k+1$.
But if $\mathsf{rank}\hat{F}\left(\mathbf{x}\right)=k+1$, then the
functions $F_{1},...,F_{k},G$ are independent and the equality $G=P\circ\left(F_{1},...,F_{k}\right)$
does not hold for any $P$. For the converse, use the constant rank
theorem (see for instance Ref. \cite{boo}, Theorem 7.1)$.\;\;\;\diamondsuit$
\begin{lem}
\label{lindep}Given a function $G:A\subseteq\mathbb{R}^{m}\rightarrow\mathbb{R}$
and a set of independent functions $F_{1},...,F_{k}:A\subseteq\mathbb{R}^{m}\rightarrow\mathbb{R}$,
the subset $A$ can be written as a disjoint union $A=D\cup I\cup B$
where $D$ and $I$ are open subsets, $B$ is a closed set (relative
to $A$) with empty interior, the function $G$ is locally dependent
on $F_{1},...,F_{k}$ along $D$ and $F_{1},...,F_{k},G$ are independent
along $I$.
\end{lem}
\textit{Proof}. Define 
\[
R_{j}\coloneqq\left\{ \mathbf{x}\in A:\mathsf{rank}\hat{F}\left(\mathbf{x}\right)=j\right\} ,
\]
with $\hat{F}$ as in the previous lemma. It is clear that $R_{j}=\emptyset$
if $j\neq k,k+1$. Accordingly, $A=R_{k}\cup R_{k+1}$. Of course,
$R_{k}=A-R_{k+1}$ (i.e. $R_{k}\cap R_{k+1}=\emptyset$) and, since
$R_{k+1}$ is open (because $k+1$ is the maximal rank), then $R_{k}$
is closed inside $A$ and we can write $R_{k}=\mathsf{int}R_{k}\cup\partial R_{k}$
(here ``$\mathsf{int}$'' and ``$\partial$'' are the interior
and the border relative to $A$). Thus, the lemma follows by taking
\[
D\coloneqq\mathsf{int}R_{k},\;\;\;I\coloneqq R_{k+1}\;\;\;\textrm{and}\;\;\;B\coloneqq\partial R_{k},
\]
and using in $D$ the constant rank theorem$.\;\;\;\diamondsuit$
\begin{lem}
\label{lema1}Given a set of independent functions $F_{1},...,F_{l}$,
and defining $F\coloneqq\left(F_{1},...,F_{l}\right)$, the following
statements are equivalent:
\end{lem}
\begin{enumerate}
\item $F_{1},...,F_{l}$ are isotropic;
\item the function $F$ satisfies
\begin{equation}
\dim\left[\mathsf{Ker}\left[DF\cdot J\cdot\left(DF\right)^{t}\right]\right]=2n-l;\label{isot}
\end{equation}
\item the function $F$ satisfies
\begin{equation}
\mathsf{Ker}\left[DF\right]\subseteq\mathsf{Im}\left[J\cdot\left(DF\right)^{t}\right].\label{sinc}
\end{equation}
\end{enumerate}
\textit{Proof}. It is easy to show that
\[
\left\{ F_{i},F_{j}\right\} =\left[DF\cdot J\cdot\left(DF\right)^{t}\right]_{ij},
\]
so the equivalence between $1$ and $2$ is immediate. Now, let us
show the equivalence between $2$ and $3$. Note that, since $DF$
is surjective (because the functions $F_{i}$'s are independent),
its transpose $\left(DF\right)^{t}$ is injective and, consequently,
since $J$ is injective too, we have that 
\begin{equation}
\begin{array}{lll}
\dim\left[\mathsf{Ker}\left[DF\cdot J\cdot\left(DF\right)^{t}\right]\right] & = & \dim\left[\left(J\cdot\left(DF\right)^{t}\right)^{-1}\cdot\mathsf{Ker}\left[DF\right]\right]\\
\\
 & = & \dim\left[\mathsf{Im}\left[J\cdot\left(DF\right)^{t}\right]\cap\mathsf{Ker}\left[DF\right]\right].
\end{array}\label{tres}
\end{equation}
On the other hand, the surjectivity of $DF$ also says that $\dim\left[\mathsf{Ker}\left[DF\right]\right]=2n-l$.
So, using \eqref{tres}, it follows that \eqref{isot} holds if and
only if \eqref{sinc} holds$.\;\;\;\diamondsuit$
\begin{rem*}
Given a linear space $V$ and a subspace $U\subseteq V$, the annihilator\textbf{
}of $U$ (i.e. the linear forms vanishing on $U$) will be denoted
$U^{0}$.
\end{rem*}
\begin{lem}
\label{lisot}Given a set of independent isotropic functions $F_{1},...,F_{k}:A\subseteq\mathbb{R}^{m}\rightarrow\mathbb{R}$,
if we add a new function $F_{k+1}:A\subseteq\mathbb{R}^{m}\rightarrow\mathbb{R}$
such that $F_{1},...,F_{k},F_{k+1}$ is an independent set, then such
a bigger set is also isotropic.
\end{lem}
\textit{Proof}. Define $F\coloneqq\left(F_{1},...,F_{k}\right)$ and
$\hat{F}\coloneqq\left(F_{1},...,F_{k},F_{k+1}\right)$. According
to Lemma \ref{lema1}, the isotropy condition on $F$ is equivalent
to Eq. \eqref{sinc}, which in turn can be written as
\[
\mathsf{Ker}\left[DF\right]\subseteq J\cdot\left(\mathsf{Ker}\left[DF\right]\right)^{0}.
\]
Here, we are using that
\begin{equation}
\mathsf{Ker}\left[DF\left(\mathbf{x}\right)\right]=\left(\mathsf{Im}\left[\left(DF\left(\mathbf{x}\right)\right)^{t}\right]\right)^{0}.\label{kim}
\end{equation}
It is clear that $\mathsf{Ker}\left[D\hat{F}\right]\subseteq\mathsf{Ker}\left[DF\right]$.
Consequently,
\[
\mathsf{Ker}\left[D\hat{F}\right]\subseteq J\cdot\left(\mathsf{Ker}\left[D\hat{F}\right]\right)^{0},
\]
what implies (by Lemma \ref{lema1} again) that $\hat{F}$ is isotropic$.\;\;\;\diamondsuit$

\bigskip{}
Now, the annunciated construction.
\begin{prop}
\label{fitn}Suppose that $F_{1},...,F_{k}:A\subseteq\mathbb{R}^{m}\rightarrow\mathbb{R}$
are functions satisfying (1), (2) and (3). Then, around every point
of an open dense subset of $A$, a set of functions $F_{k+1},...,F_{l}$
can be constructed such that $F_{1},...,F_{l}$ satisfy the conditions
(1), (2), (3) and (4).
\end{prop}
\textit{Proof}. Let us consider, for each $r\in\mathbb{N}$, the set
of pairs
\[
S_{r}\coloneqq\left\{ \left(i,j\right):i,j\in\left\{ 1,...,r\right\} ,i<j\right\} 
\]
and a bijection $\phi_{r}:S_{r}\rightarrow\left\{ 1,...,r\,\left(r-1\right)\right\} $.
(This is just to simplify the notation). Given $\left(i,j\right)\in S_{k}$
with $\phi_{k}\left(i,j\right)=a$, consider the set of first integrals
$F_{1},...,F_{k},G^{a}$, where $G^{a}\coloneqq\left\{ F_{i},F_{j}\right\} $
(recall that the Poisson bracket of two first integrals is also a
first integral). According to Lemma \ref{lindep}, we can decompose
$A$ as a disjoint union $A=D^{a}\cup I^{a}\cup B^{a}$ with the properties
mentioned in such a lemma. It is clear that, in the open subset $\hat{D}\coloneqq\bigcap_{a=1}^{k\left(k-1\right)}D^{a}$,
the condition (4) is locally fulfilled for the functions $F_{1},...,F_{k}$.
Then, around every point of $\hat{D}$, conditions (1), (2), (3) and
(4) are true for $F_{1},...,F_{k}$. Note that the complement of $\hat{D}$
(inside $A$) is given by the union
\[
\bigcup_{a=1}^{k\left(k-1\right)}\left(I^{a}\cup B^{a}\right).
\]
Then, we can write
\[
A=\hat{D}\cup\left(\cup_{a}I^{a}\right)\cup\hat{B},
\]
where $\hat{B}\coloneqq\bigcup_{a=1}^{k\left(k-1\right)}B^{a}$ is
a closed set (relative to $A$) with empty interior. 

Let us focus on each open subset $I^{a}$. There, the functions $F_{1},...,F_{k},F_{k+1}\coloneqq G^{a}$
satisfy (1) by definition of $I^{a}$, (2) because they are first
integrals, and, according to Lemma \ref{lisot}, they also satisfy
(3). Now, given $\left(i',j'\right)\in S_{k+1}$ with $\phi_{k+1}\left(i',j'\right)=b$,
consider the set of first integrals $F_{1},...,F_{k},F_{k+1},G^{b}$,
with $G^{b}=\left\{ F_{i'},F_{j'}\right\} $, and the decomposition
$I^{a}=D^{a,b}\cup I^{a,b}\cup B^{a,b}$ of Lemma \ref{lindep}. Again,
around every point of each open subset $\hat{D}^{a}=\bigcap_{b=1}^{\left(k+1\right)k}D^{a,b}$,
the conditions (1), (2), (3) and (4) are true for the function $F_{1},...,F_{k},F_{k+1}$.
And we can write
\[
I^{a}=\hat{D}^{a}\cup\left(\cup_{b}I^{a,b}\right)\cup\hat{B}^{a},
\]
where each $\hat{B}^{a}\coloneqq\bigcup_{b=1}^{\left(k+1\right)k}B^{a,b}$
is a closed set (relative to $I^{a}$) with empty interior. Applying
this procedure $s$ times, we shall arrive at:
\begin{itemize}
\item open subsets $I^{a_{1},...,a_{s}}$, contained in $I^{a_{1},...,a_{s-1}}$,
where the functions
\[
F_{1},...,F_{k},F_{k+1}\coloneqq G^{a_{1}},F_{k+2}\coloneqq G^{a_{2}},...,F_{k+s}\coloneqq G^{a_{s}}
\]
satisfy (1), (2) and (3);
\item open subsets $\hat{D}^{a_{1},...,a_{s-1}}$ where
\[
F_{1},...,F_{k},F_{k+1}=G^{a_{1}},F_{k+2}=G^{a_{2}},...,F_{k+s-1}=G^{a_{s-1}}
\]
satisfy (1), (2), (3) and (4) around each point of it;
\item and closed sets (relative to $I^{a_{1},...,a_{s-1}}$) with empty
interior $\hat{B}^{a_{1},...,a_{s-1}}$ such that
\[
I^{a_{1},...,a_{s-1}}=\hat{D}^{a_{1},...,a_{s-1}}\cup\left(\cup_{a_{s}}I^{a_{1},...,a_{s}}\right)\cup\hat{B}^{a_{1},...,a_{s-1}}.
\]
\end{itemize}
When $s=m-k$, we shall have open subsets $I^{a_{1},a_{2},...,a_{m-k}}$
where 
\[
F_{1},...,F_{k},F_{k+1}=G^{a_{1}},F_{k+2}=G^{a_{2}},...,F_{m}=G^{a_{m-k}}
\]
are independent, i.e. $F=\left(F_{1},...,F_{m}\right)$ has rank $m$
there, which is the maximal one. Then, in the next step,
\[
I^{a_{1},...,a_{m-k},b}=B^{a_{1},...,a_{m-k},b}=\emptyset
\]
 for all $b$, so $I^{a_{1},...,a_{m-k}}=\hat{D}^{a_{1},...,a_{m-k}}$.
As a consequence, we can write $A=D\cup B$ where
\[
D=\hat{D}\cup\left(\cup_{a_{1}}\hat{D}^{a_{1}}\right)\cup\left(\cup_{a_{1},a_{2}}\hat{D}^{a_{1},a_{2}}\right)\cup\cdots\cup\left(\cup_{a_{1},...,a_{m-k}}\hat{D}^{a_{1},...,a_{m-k}}\right)
\]
is an open subset such that, around every point of it, functions $F_{k+1},...,F_{l}$
can be constructed for which $F_{1},...,F_{l}$ satisfy conditions
(1), (2), (3) and (4), and 
\[
B=\hat{B}\cup\left(\cup_{a_{1}}\hat{B}^{a_{1}}\right)\cup\left(\cup_{a_{1},a_{2}}\hat{B}^{a_{1},a_{2}}\right)\cup\cdots\cup\left(\cup_{a_{1},...,a_{m-k-1}}\hat{B}^{a_{1},...,a_{m-k-1}}\right)
\]
is a set with empty interior. The last fact says that $D$ is dense
inside $A$, what ends our proof$.\;\;\;\diamondsuit$

\bigskip{}

Combining Theorem \ref{ncies} and the last proposition, we easily
have that:
\begin{thm}
\label{dens} Given a Hamiltonian system, if we know a set of isotropic
and independent first integrals, then the system is exactly solvable
along an open dense subset of the phase space.
\end{thm}
\smallskip{}

\begin{rem}
\label{cjl} Given a Hamiltonian system, the existence of a set of
local isotropic first integrals can be easily shown as a direct consequence
of the Carathéodory-Jacobi-Lie theorem (see the Appendix). But in
the theorem above (as in the definition of a NCI system) it is not
the existence of a set of isotropic first integrals what is required,
but the knowledge of such a set. Otherwise, the construction of the
trajectories can not be done.
\end{rem}
\smallskip{}

\begin{rem}
\label{nonex} Theorem \ref{dens} can be seen as a new criterium
(i.e. a sufficient condition) for exact solvability of Hamiltonian
systems, which is weaker than the NCI. Its usefulness is very clear:
if for a given Hamiltonian system we know a set of isotropic first
integrals (which do not necessarily satisfy the closure condition),
then we can be sure that the system is exactly solvable. But such
a new criterium does not give rise to new exactly solvable systems,
beyond the NCI ones, as it is clear from Proposition \ref{fitn}.
In other words, it is not possible to give examples of Hamiltonian
systems satisfying this new criterium which are not (in essence) NCI
systems.
\end{rem}
In this way, we have given our first proof to the main result of the
paper. It says that the closure condition is not essential, a priori,
for ensuring exact solvability. Nevertheless, if we go over the above
results, we can see that, in order to construct the trajectories,
we previously need to construct more first integrals, in such a way
that the resulting entire set of first integrals gives rise to a NCI
system. Thus, closure condition is involved in the integration process
at the end of the day. This does not contradict our result, it simply
says that in its proof such a condition still plays an important role.
In the last section of the paper we shall present an alternative procedure
for integrating the Hamilton equations that only uses the isotropic
first integrals. In other words, we give a second proof of our result
in which the closure condition is not used at all.

\section{The standard Hamilton-Jacobi theory}

The main idea behind the Hamilton-Jacobi theory is to find coordinates
where the equations of motion of a Hamiltonian system adopt a very
simple form \cite{ar,gold}. Let us review such an idea. 

\bigskip{}

Consider a Hamiltonian system with $n$ degrees of freedom defined
by a Hamiltonian function $H$. The (time independent) \textbf{Hamilton-Jacobi
equation (HJE)} is 
\begin{equation}
\frac{\partial}{\partial\mathbf{q}}\left[H\left(\mathbf{q},\frac{\partial W}{\partial\mathbf{q}}\left(\mathbf{q}\right)\right)\right]=0,\label{class}
\end{equation}
whose unknown is a function $W:\mathbb{R}^{n}\rightarrow\mathbb{R}$.
The solutions $W$ of such an equation are called\textit{ Hamilton's
characteristic functions}. 

In practice, $W$ is usually defined only along an open subset of
$\mathbb{R}^{n}$. In such a case, one says that $W$ is a \textbf{local
solution}. Nevertheless, to simplify the notation, we shall assume
that the domain is always the entire space.

One is actually interested in finding a ``big enough'' family of
such solutions or, more precisely, a function $W:\mathbb{R}^{n}\times\mathbb{R}^{n}\rightarrow\mathbb{R}$
such that each
\[
W_{\lambda}\coloneqq W\left(\cdot,\lambda\right):\mathbb{R}^{n}\rightarrow\mathbb{R}:\mathbf{q}\mapsto W\left(\mathbf{q},\mathbf{\lambda}\right),\;\;\;\lambda\in\mathbb{R}^{n},
\]
 is a solution of the HJE and 
\begin{equation}
\det\left[\frac{\partial^{2}W}{\partial\mathbf{\lambda}\partial\mathbf{q}}\left(\mathbf{q},\lambda\right)\right]\neq0,\;\;\;\forall\left(\mathbf{q},\lambda\right)\in\mathbb{R}^{2n}.\label{det}
\end{equation}
Each function $W_{\lambda}$ is called a \textbf{partial solution}
of the HJE. 
\begin{rem*}
Note that the HJE implies that $H\left(\mathbf{q},\frac{\partial W}{\partial\mathbf{q}}\left(\mathbf{q},\lambda\right)\right)$
only depends on $\mathbf{\lambda}$, i.e. 
\begin{equation}
H\left(\mathbf{q},\frac{\partial W}{\partial\mathbf{q}}\left(\mathbf{q},\lambda\right)\right)=h\left(\mathbf{\lambda}\right)\label{hl}
\end{equation}
for some function $h:\mathbb{R}^{n}\rightarrow\mathbb{R}$.
\end{rem*}
The condition \eqref{det} is the same as asking that the function
$\Sigma:\mathbb{R}^{n}\times\mathbb{R}^{n}\rightarrow\mathbb{R}^{2n},$
given by
\begin{equation}
\Sigma\left(\mathbf{q},\lambda\right)=\left(\mathbf{q},\frac{\partial W}{\partial\mathbf{q}}\left(\mathbf{q},\lambda\right)\right),\label{Sql}
\end{equation}
is a \textit{local} diffeomorphism. This means that, for every couple
of points $\left(\mathbf{q},\lambda\right),\left(\mathbf{q},\mathbf{p}\right)\in\mathbb{R}^{2n}$
such that $\Sigma\left(\mathbf{q},\lambda\right)=\left(\mathbf{q},\mathbf{p}\right)$,
there exist open neighborhoods $V$ and $U$ of $\left(\mathbf{q},\lambda\right)$
and $\left(\mathbf{q},\mathbf{p}\right)$, respectively, such that
$\Sigma\left(V\right)=U$ and the restriction of $\Sigma$ to $V$
is a diffeomorphism with its image $U$. 
\begin{rem*}
Unless a confusion may arise, every local inverse of $\Sigma$ will
be indicated simply as $\Sigma^{-1}$ (no mention to the domain or
codomain). 
\end{rem*}
We shall also ask $\Sigma$ to be surjective. A function $\Sigma$
{[}given by \eqref{Sql}{]} with all these properties is called a
\textbf{complete solution} of the HJE. If $\Sigma$ is not defined
along all of $\mathbb{R}^{n}\times\mathbb{R}^{n}$, but along an open
subset of it, we shall say that $\Sigma$ is a \textbf{local complete
solution}.

\bigskip{}

Given a complete solution $\Sigma$, it can be shown \cite{ar,gold}
that the equations
\begin{equation}
\mathbf{Q}=\frac{\partial W}{\partial\lambda}\left(\mathbf{q},\lambda\right),\;\;\;\mathbf{p}=\frac{\partial W}{\partial\mathbf{q}}\left(\mathbf{q},\lambda\right),\label{Qp}
\end{equation}
define a new set of (local) canonical coordinates $\mathbf{Q}=\mathbf{Q}\left(\mathbf{q},\mathbf{p}\right)$
and $\lambda=\lambda\left(\mathbf{q},\mathbf{p}\right)$ around every
point of the phase space, in terms of which the canonical Hamilton
equations {[}recall \eqref{ceq}{]} read
\[
\dot{\mathbf{Q}}\left(t\right)=\frac{\partial K}{\partial\mathbf{\lambda}}\left(\mathbf{Q}\left(t\right),\mathbf{\lambda}\left(t\right)\right),\;\;\;\dot{\mathbf{\lambda}}\left(t\right)=-\frac{\partial K}{\partial\mathbf{Q}}\left(\mathbf{Q}\left(t\right),\mathbf{\lambda}\left(t\right)\right),
\]
where $K\left(\mathbf{Q},\mathbf{\lambda}\right)=H\left(\mathbf{q}\left(\mathbf{Q},\mathbf{\lambda}\right),\mathbf{p}\left(\mathbf{Q},\mathbf{\lambda}\right)\right)$.
The crucial point here is that {[}see Eq. \eqref{hl}{]}
\[
K\left(\mathbf{Q},\mathbf{\lambda}\right)=H\left(\mathbf{q}\left(\mathbf{Q},\mathbf{\lambda}\right),\frac{\partial W}{\partial\mathbf{q}}\left(\mathbf{q}\left(\mathbf{Q},\mathbf{\lambda}\right),\mathbf{\lambda}\right)\right)=h\left(\mathbf{\lambda}\right),
\]
and consequently the equations of motion translate to
\[
\dot{\mathbf{Q}}\left(t\right)=\frac{\partial h}{\partial\mathbf{\lambda}}\left(\mathbf{\lambda}\left(t\right)\right),\;\;\;\dot{\mathbf{\lambda}}\left(t\right)=0,
\]
which can be easily solved. In fact, the general solution is given
by
\[
\mathbf{Q}\left(t\right)=\mathbf{Q}_{0}+t\,\frac{\partial h}{\partial\mathbf{\lambda}}\left(\mathbf{\lambda}_{0}\right),\;\;\;\lambda\left(t\right)=\lambda_{0}.
\]
Moreover, the trajectories $\left(\mathbf{q}\left(t\right),\mathbf{p}\left(t\right)\right)$
of the system can be obtained through the algebraic equations {[}see
\eqref{Qp}{]}
\[
\frac{\partial W}{\partial\mathbf{\lambda}}\left(\mathbf{q}\left(t\right),\lambda_{0}\right)=\mathbf{Q}_{0}+t\,\frac{\partial h}{\partial\mathbf{\lambda}}\left(\mathbf{\lambda}_{0}\right)\;\;\;\textrm{and}\;\;\;\mathbf{p}\left(t\right)=\frac{\partial W}{\partial\mathbf{q}}\left(\mathbf{q}\left(t\right),\lambda_{0}\right).
\]
We just must solve the first equation for $\mathbf{q}\left(t\right)$,
which can be done because of condition \eqref{det}.

On the other hand, the functions $F_{i}\left(\mathbf{q},\mathbf{p}\right)\coloneqq\lambda_{i}\left(\mathbf{q},\mathbf{p}\right)$
are local first integrals of the system and, since they are conjugate
momenta, they are in involution, i.e. $\left\{ F_{i},F_{j}\right\} =0$
for all $i,j=1,...,n$. Summing up,
\begin{thm}
\label{shjt}Consider a Hamiltonian system with $n$ degrees of freedom.
If we know a complete solution $\Sigma$ of the HJE for such a system,
then the latter can be exactly solved. Moreover, the system is locally
commutative integrable by means of the local first integrals
\[
F_{i}\left(\mathbf{q},\mathbf{p}\right)=\lambda_{i}\left(\mathbf{q},\mathbf{p}\right)=\left[\Sigma^{-1}\left(\mathbf{q},\mathbf{p}\right)\right]_{n+i},\;\;\;i=1,...,n.
\]
\end{thm}
The second affirmation in the last theorem establishes a deep connection
between commutative integrability and the Hamilton-Jacobi theory:
\textit{given a complete solution of the HJE, we have, around every
point of $\mathbb{R}^{2n}$, $n$ local independent first integrals
in involution}. A reciprocal result is also true, under an additional
assumption. Suppose that we have a set of $n$ functions $F_{1},...,F_{n}$
such that the set of vectors
\[
\left\{ \frac{\partial F_{1}}{\partial\mathbf{p}}\left(\mathbf{q},\mathbf{p}\right),...,\frac{\partial F_{n}}{\partial\mathbf{p}}\left(\mathbf{q},\mathbf{p}\right)\right\} 
\]
is l.i. for all $\left(\mathbf{q},\mathbf{p}\right)$. One says that
the functions $F_{1},...,F_{n}$ are \textit{vertically independent.}
(This implies, in particular, that the involved functions are independent).
Now, denote $\pi:\mathbb{R}^{2n}\rightarrow\mathbb{R}^{n}$ the projection
$\pi\left(\mathbf{q},\mathbf{p}\right)=\mathbf{q}$ and define $F\coloneqq\left(F_{1},...,F_{n}\right):\mathbb{R}^{2n}\rightarrow\mathbb{R}^{n}$.
It can be shown that $\left(\pi,F\right):\mathbb{R}^{2n}\rightarrow\mathbb{R}^{n}\times\mathbb{R}^{n}$
is a local diffeomorphism. Moreover, if the functions $F_{i}$ are
first integrals for a Hamiltonian $H$ and they are in involution,
then each inverse $\Sigma\coloneqq\left(\pi,F\right)^{-1}$ is a (local)
complete solution of the HJE for $H$. Thus: \textit{given $n$ vertically
independent first integrals in involution, we have, around every point
of of $\mathbb{R}^{2n}$, a local complete solution of the HJE}.

\bigskip{}

All above results will be shown in the next sections, in a more general
context.

\section{An extended Hamilton-Jacobi theory}

We have said in the last section that there is a deep connection between
the (standard) Hamilton-Jacobi theory and the commutative integrability.
Based on Ref. \cite{gp}, we shall present below a slightly extension
of such a theory which is intimately related to the non-commutative
integrability. (Another extension of the Hamilton-Jacobi theory related
to NCI has been developed in \cite{koz}).

\subsection{Re-writing the HJE}

Fix a Hamiltonian function $H:\mathbb{R}^{2n}\rightarrow\mathbb{R}$.
From now on, given a function $F:A\subseteq\mathbb{R}^{m}\rightarrow\mathbb{R}$,
by $\nabla F\left(\mathbf{x}\right)$ we shall denote the column vector
$\left(DF\left(\mathbf{x}\right)\right)^{t}$. Given a solution $W$
of \eqref{class}, let us define $\sigma:\mathbb{R}^{n}\rightarrow\mathbb{R}^{2n}$
as
\[
\sigma\left(\mathbf{q}\right)\coloneqq\left(\mathbf{q},\hat{\sigma}\left(\mathbf{q}\right)\right)\coloneqq\left(\mathbf{q},\left(\nabla W\left(\mathbf{q}\right)\right)^{t}\right).
\]
It is clear that $\sigma$ satisfies
\begin{equation}
\nabla\left(H\circ\sigma\right)=0\;\;\;\textrm{and}\;\;\;\nabla\times\hat{\sigma}^{t}=0.\label{of}
\end{equation}
Reciprocally, given $\sigma:\mathbb{R}^{n}\rightarrow\mathbb{R}^{2n}$
of the form 
\begin{equation}
\sigma\left(\mathbf{q}\right)\coloneqq\left(\mathbf{q},\hat{\sigma}\left(\mathbf{q}\right)\right)\label{form}
\end{equation}
and fulfilling \eqref{of}, then $\left(\hat{\sigma}\left(\mathbf{q}\right)\right)^{t}=\nabla W\left(\mathbf{q}\right)$
for some function $W$ satisfying the HJE. So, we can think of \eqref{of}
as the HJE and take the functions $\sigma$ of the form \eqref{form}
as their unknowns. In these terms, the complete solutions will be
given by a family of solutions $\sigma_{\lambda}$ such that $\Sigma\left(\mathbf{q},\lambda\right)=\sigma_{\lambda}\left(\mathbf{q}\right)$
is a surjective local diffeomorphism. But we shall consider a further
modification of \eqref{of}. 
\begin{rem}
\label{lefti} Note that a function of the form \eqref{form} has
a left inverse (of class $C^{\infty}$), and consequently the same
is true for each differential $D\sigma\left(\mathbf{q}\right)$. One
of its left inverses is the projection $\pi:\mathbb{R}^{2n}\rightarrow\mathbb{R}^{n}:\left(\mathbf{q},\mathbf{p}\right)\mapsto\mathbf{q}$.
In particular, $D\sigma\left(\mathbf{q}\right):\mathbb{R}^{n}\rightarrow\mathbb{R}^{2n}$
is injective for all $\mathbf{q}$, what means that 
\begin{equation}
\dim\left[\mathsf{Im}\left[D\sigma\left(\mathbf{q}\right)\right]\right]=n,\;\;\;\forall\mathbf{q}\in\mathbb{R}^{n}.\label{dim}
\end{equation}
Fix a solution $\sigma$ of \eqref{of}. On the one hand, since
\[
\nabla\left(H\circ\sigma\right)\left(\mathbf{q}\right)=\left(D\sigma\left(\mathbf{q}\right)\right)^{t}\cdot\left(\nabla H\circ\sigma\left(\mathbf{q}\right)\right),
\]
we have that 
\begin{equation}
\nabla H\circ\sigma\left(\mathbf{q}\right)\in\mathsf{Ker}\left[\left(D\sigma\right)^{t}\left(\mathbf{q}\right)\right],\;\;\;\forall\mathbf{q}\in\mathbb{R}^{n}.\label{eq:1}
\end{equation}
On the other hand, in terms of the matrix $J$ {[}see Eq. \eqref{J}{]},
the condition $\nabla\times\hat{\sigma}^{t}=0$ is equivalent to
\begin{equation}
\left(D\sigma\left(\mathbf{q}\right)\right)^{t}\cdot J\cdot D\sigma\left(\mathbf{q}\right)=0,\label{rot}
\end{equation}
what says that 
\[
\mathsf{Im}\left[J\cdot D\sigma\left(\mathbf{q}\right)\right]\subseteq\mathsf{Ker}\left[\left(D\sigma\left(\mathbf{q}\right)\right)^{t}\right].
\]
Since {[}recall Eq. \eqref{kim}{]}
\[
\mathsf{Ker}\left[\left(D\sigma\left(\mathbf{q}\right)\right)^{t}\right]=\left(\mathsf{Im}\left[D\sigma\left(\mathbf{q}\right)\right]\right){}^{0},
\]
then {[}see \eqref{dim}{]}
\begin{equation}
\dim\left[\mathsf{Ker}\left[\left(D\sigma\right)^{t}\left(\mathbf{q}\right)\right]\right]=2n-\dim\left[\mathsf{Im}\left[D\sigma\left(\mathbf{q}\right)\right]\right]=2n-n=n.\label{n}
\end{equation}
Accordingly, using that $\dim\left[\mathsf{Im}\left[J\cdot D\sigma\left(\mathbf{q}\right)\right]\right]=\dim\left[\mathsf{Im}\left[D\sigma\left(\mathbf{q}\right)\right]\right]$
(since $J$ is invertible), we have the equality
\begin{equation}
\mathsf{Ker}\left[\left(D\sigma\right)^{t}\left(\mathbf{q}\right)\right]=\mathsf{Im}\left[J\cdot D\sigma\left(\mathbf{q}\right)\right].\label{eq:2}
\end{equation}
\end{rem}
So, combining Eqs. \eqref{eq:1} and \eqref{eq:2},
\begin{equation}
\nabla H\circ\sigma\left(\mathbf{q}\right)\in\mathsf{Im}\left[J\cdot D\sigma\left(\mathbf{q}\right)\right],\;\;\;\forall\mathbf{q}\in\mathbb{R}^{n}.\label{fin}
\end{equation}
Concluding,
\begin{prop}
A function of the form \eqref{form} satisfies \eqref{of} if and
only if satisfies {[}see \eqref{rot} and \eqref{fin}{]}
\begin{equation}
\nabla H\circ\sigma\left(\mathbf{q}\right)\in J\cdot\mathsf{Im}\left(D\sigma\left(\mathbf{q}\right)\right)\;\;\;\textrm{and}\;\;\;\left(D\sigma\left(\mathbf{q}\right)\right)^{t}\cdot J\cdot\left(D\sigma\left(\mathbf{q}\right)\right)=0,\label{bf}
\end{equation}
for all $\mathbf{q}\in\mathbb{R}^{n}$.
\end{prop}
\begin{rem*}
The first part of \eqref{bf} is important from the geometric point
of view, because it says that the image of $\sigma$ defines an invariant
manifold for the Hamiltonian system. The second part says that such
a manifold is \textit{Lagrangian} \cite{am,mr}.
\end{rem*}

\subsection{Generalized solutions of the HJE}

Now, the annunciated extension.
\begin{defn}
A \textbf{generalized (partial) solution} of the HJE for $H$ is a
left invertible function $\sigma:\mathbb{R}^{r}\rightarrow\mathbb{R}^{2n}$
(see Remark \ref{lefti}), for some natural $r$, satisfying {[}see
Eq. \eqref{bf}{]}
\begin{equation}
\nabla H\circ\sigma\left(\mathbf{x}\right)\in J\cdot\mathsf{Im}\left(D\sigma\left(\mathbf{x}\right)\right)\;\;\;\textrm{and}\;\;\;\left(D\sigma\left(\mathbf{x}\right)\right)^{t}\cdot J\cdot\left(D\sigma\left(\mathbf{x}\right)\right)=0,\label{bf-1}
\end{equation}
 for all $\mathbf{x}\in\mathbb{R}^{r}$. And a \textbf{generalized
complete solution} of the HJE for $H$ is a family of partial solutions
$\sigma_{\lambda}:\mathbb{R}^{r}\rightarrow\mathbb{R}^{2n}$ with
the same left inverse, and with $\lambda\in\mathbb{R}^{l}$ for some
natural $l$, such that 
\[
\Sigma:\mathbb{R}^{r}\times\mathbb{R}^{l}\rightarrow\mathbb{R}^{2n}:\left(\mathbf{x},\lambda\right)\mapsto\sigma_{\lambda}\left(\mathbf{x}\right)
\]
 is a surjective local diffeomorphism.

When the function $\sigma$ (resp. $\Sigma$) is defined along a proper
open subset of $\mathbb{R}^{r}$ (resp. $\mathbb{R}^{r}\times\mathbb{R}^{l}$),
we shall say that $\sigma$ (resp. $\Sigma$) is a \textbf{generalized
local solution} (resp. \textbf{generalized local complete solution})
of the HJE. 
\end{defn}
\begin{rem*}
In Ref. \cite{gp}, we have called ``solutions'' to the functions
$\sigma$ satisfying just the first part of \eqref{bf-1}, and ``isotropic
solutions'' to those $\sigma$ that also satisfy the second part.
This is because the image of $\sigma$ defines an \textit{isotropic}
\cite{am,mr} invariant manifold.
\end{rem*}
The values of $r$ and $l$ are not arbitrary.
\begin{prop}
If $\sigma:\mathbb{R}^{r}\rightarrow\mathbb{R}^{2n}$ is a generalized
solution of the HJE for $H$, then $r\leq n$. And if $\Sigma:\mathbb{R}^{r}\times\mathbb{R}^{l}\rightarrow\mathbb{R}^{2n}$
is a generalized complete solution, then $r+l=2n$.
\end{prop}
\textit{Proof}. The second part of \eqref{bf-1} is equivalent to
\begin{equation}
\mathsf{Im}\left[J\cdot D\sigma\left(\mathbf{x}\right)\right]\subseteq\mathsf{Ker}\left[\left(D\sigma\left(\mathbf{x}\right)\right)^{t}\right].\label{inc}
\end{equation}
Since $\sigma$ is left invertible, then 
\begin{equation}
\dim\left[\mathsf{Im}\left[D\sigma\left(\mathbf{x}\right)\right]\right]=r,\label{dsr}
\end{equation}
and consequently 
\begin{equation}
\dim\left[\mathsf{Ker}\left[\left(D\sigma\left(\mathbf{x}\right)\right)^{t}\right]\right]=2n-\dim\left[\mathsf{Im}\left[D\sigma\left(\mathbf{x}\right)\right]\right]=2n-r.\label{r}
\end{equation}
Finally, since $J$ is invertible, $\dim\left[\mathsf{Im}\left[J\cdot D\sigma\left(\mathbf{x}\right)\right]\right]=\dim\left[\mathsf{Im}\left[D\sigma\left(\mathbf{x}\right)\right]\right]$.
So, from the last three equations we have that $r\leq2n-r$, or equivalently,
$r\leq n$. 

Regarding the second affirmation of the proposition, the fact that
$\Sigma$ must be a local diffeomorphism implies that $l+r=2n$ (see
Ref. \cite{boo})$.\;\;\;\diamondsuit$
\begin{rem*}
Consider a function $\sigma$ satisfying the second part of \eqref{bf-1}.
If $\sigma$ also satisfies the first part, it is clear that 
\begin{equation}
\nabla\left(H\circ\sigma\right)\left(\mathbf{x}\right)=0,\;\;\;\forall\mathbf{x}\in\mathbb{R}^{r}.\label{gradH}
\end{equation}
But the converse is not true {[}i.e. if $\sigma$ satisfies \eqref{gradH},
it is not true, in general, that it satisfies the first part of \eqref{bf-1}{]},
as happens for the $r=n$ case. This is precisely because the inclusion
\eqref{inc} is strict for $r<n$ {[}compare \eqref{n} and \eqref{r}{]}.
\end{rem*}
The next result gives an alternative way of describing the generalized
partial and complete solutions of the HJE which will be useful later.
\begin{prop}
A left invertible function $\sigma$ satisfying the second part of
\eqref{bf-1} is a generalized solution of the HJE for $H$ if and
only if 
\[
\nabla H\circ\sigma\left(\mathbf{x}\right)=J\cdot D\sigma\left(\mathbf{x}\right)\cdot v_{H}^{\sigma}\left(\mathbf{x}\right),
\]
with
\begin{equation}
v_{H}^{\sigma}\left(\mathbf{x}\right)\coloneqq-\left[D\Pi\left(\sigma\left(\mathbf{x}\right)\right)\right]\cdot J\cdot\left(\nabla H\circ\sigma\left(\mathbf{x}\right)\right),\label{vhs}
\end{equation}
being $\Pi:\mathbb{R}^{2n}\rightarrow\mathbb{R}^{r}$ some left inverse
of $\sigma$. And a family of left invertible functions $\sigma_{\lambda}$,
with common left inverse $\Pi$, defines a generalized complete solution
$\Sigma$ if and only if $\Sigma$ is a surjective local diffeomorphism,
\begin{equation}
\nabla\left(H\circ\Sigma\right)\left(\mathbf{x},\lambda\right)=\left(D\Sigma\left(\mathbf{x},\lambda\right)\right)^{t}\cdot J\cdot D\Sigma\left(\mathbf{x},\lambda\right)\cdot\left(\begin{array}{c}
v_{H}^{\sigma_{\lambda}}\left(\mathbf{x}\right)\\
0
\end{array}\right)\label{mm}
\end{equation}
{[}where $v_{H}^{\sigma_{\lambda}}\left(\mathbf{x}\right)$ is given
by \eqref{vhs}{]} and
\begin{equation}
\left(\begin{array}{c}
w\\
0
\end{array}\right)^{t}\cdot\left(D\Sigma\left(\mathbf{x},\lambda\right)\right)^{t}\cdot J\cdot D\Sigma\left(\mathbf{x},\lambda\right)\cdot\left(\begin{array}{c}
v\\
0
\end{array}\right)=0,\label{iwv}
\end{equation}
for all column vectors $v,w\in\mathbb{R}^{r}$.
\end{prop}
\textit{Proof}. The condition $\nabla H\circ\sigma\left(\mathbf{x}\right)\in J\cdot\mathsf{Im}\left(D\sigma\left(\mathbf{x}\right)\right)$
is equivalent to the existence of a column vector $v\in\mathbb{R}^{r}$
such that
\[
\nabla H\circ\sigma\left(\mathbf{x}\right)=J\cdot D\sigma\left(\mathbf{x}\right)\cdot v.
\]
Let $\Pi$ be a left inverse of $\sigma$, i.e. $\Pi\circ\sigma=\mathsf{id}_{\mathbb{R}^{r}}$.
Note that $\left[D\Pi\left(\sigma\left(\mathbf{x}\right)\right)\right]\cdot D\sigma\left(\mathbf{x}\right)=I_{r}$.
Multiplying by $\left[D\Pi\left(\sigma\left(\mathbf{x}\right)\right)\right]\cdot J$
to the left both members of above equation, and using that $J^{-1}=-J$,
we have 
\[
\left[D\Pi\left(\sigma\left(\mathbf{x}\right)\right)\right]\cdot J\cdot\left(\nabla H\circ\sigma\left(\mathbf{x}\right)\right)=-v,
\]
which proves the first part of the proposition. Now, consider a complete
solution $\Sigma$ given by functions $\sigma_{\lambda}$ (all of
them with the same left inverse $\Pi$). Then, according to the last
result, they must satisfy
\begin{equation}
\nabla H\circ\sigma_{\lambda}\left(\mathbf{x}\right)=J\cdot D\sigma_{\lambda}\left(\mathbf{x}\right)\cdot v_{H}^{\sigma_{\lambda}}\left(\mathbf{x}\right).\label{m}
\end{equation}
Since 
\[
D\sigma_{\lambda}\left(\mathbf{x}\right)\cdot v_{H}^{\sigma_{\lambda}}\left(\mathbf{x}\right)=D\Sigma\left(\mathbf{x},\lambda\right)\cdot\left(\begin{array}{c}
v_{H}^{\sigma_{\lambda}}\left(\mathbf{x}\right)\\
0
\end{array}\right),
\]
multiplying \eqref{m} by $\left(D\Sigma\left(\mathbf{x},\lambda\right)\right)^{t}$
to the left, we have precisely the Eq. \eqref{mm}. To prove \eqref{iwv},
it is enough to check that 
\[
\left(D\Sigma\left(\mathbf{x},\lambda\right)\right)^{t}\cdot J\cdot D\Sigma\left(\mathbf{x},\lambda\right)\cdot\left(\begin{array}{c}
v\\
0
\end{array}\right)=\left(\begin{array}{c}
0\\
v'
\end{array}\right)
\]
for some column vector $v'$, which is true thanks to the second part
of \eqref{bf-1} (for each $\sigma_{\lambda}$). The reciprocal follows
reversing the previous steps$.\;\;\;\diamondsuit$

\section{Complete solutions and isotropic first integrals}

Now, let us see that, related to any generalized complete solution
$\Sigma$, we have a set of $l$ independent (local) first integrals.
Let 
\begin{equation}
\mathfrak{p}:\mathbb{R}^{r}\times\mathbb{R}^{l}\rightarrow\mathbb{R}^{l}:\left(\mathbf{x},\lambda\right)\mapsto\lambda\label{p}
\end{equation}
be the projection onto the second factor. Around a given point of
$\mathbb{R}^{2n}$, fix a local inverse $\Sigma^{-1}$ of $\Sigma$
and define $F=\left(F_{1},...,F_{l}\right)=\mathfrak{p}\circ\Sigma^{-1}$.
(Note that $F$ is only defined on the open subset where $\Sigma^{-1}$
is defined). Since 
\begin{equation}
DF=\left[D\mathfrak{p}\circ\Sigma^{-1}\right]\cdot\left(D\Sigma\right)^{-1},\label{dfds}
\end{equation}
$D\Sigma$ is non-singular and $D\mathfrak{p}$ is surjective, then
$DF$ is also surjective, which means that the functions $F_{i}$'s
are independent. Let us show that they are first integrals for $H$.
The Poisson bracket between each $F_{i}$ and $H$ is {[}see \eqref{pb}{]}
\[
\left\{ F_{i},H\right\} =DF_{i}\cdot J\cdot\nabla H.
\]
So, the bracket vanishes for all $i=1,...,l$ if and only if $DF\cdot J\cdot\nabla H=0$,
or equivalently,
\begin{equation}
J\cdot\nabla H\left(\mathbf{q},\mathbf{p}\right)\in\mathsf{Ker}\left[DF\left(\mathbf{q},\mathbf{p}\right)\right].\label{fi}
\end{equation}
In terms of points $\left(\mathbf{x},\lambda\right)=\Sigma^{-1}\left(\mathbf{q},\mathbf{p}\right)$,
this means that
\begin{equation}
J\cdot\nabla H\left(\Sigma\left(\mathbf{x},\lambda\right)\right)\in\mathsf{Ker}\left[DF\left(\Sigma\left(\mathbf{x},\lambda\right)\right)\right].\label{fi-1}
\end{equation}
It is easy to see that
\[
\mathsf{Ker}\left[D\mathfrak{p}\left(\mathbf{x},\lambda\right)\right]=\mathbb{R}^{r}\times\left\{ 0\right\} ,
\]
and accordingly {[}recall \eqref{dfds}{]}
\begin{equation}
\mathsf{Ker}\left[DF\left(\Sigma\left(\mathbf{x},\lambda\right)\right)\right]=\left(D\Sigma\left(\mathbf{x},\lambda\right)\right)\cdot\mathsf{Ker}\left[D\mathfrak{p}\left(\mathbf{x},\lambda\right)\right]=\mathsf{Im}\left[\frac{\partial\Sigma}{\partial\mathbf{x}}\left(\mathbf{x},\lambda\right)\right]=\mathsf{Im}\left[D\sigma_{\lambda}\left(\mathbf{x}\right)\right].\label{kerf}
\end{equation}
Thus, Eq. \eqref{fi-1} is equivalent to
\[
\nabla H\left(\Sigma\left(\mathbf{x},\lambda\right)\right)=\nabla H\circ\sigma_{\lambda}\left(\mathbf{x}\right)\in J\cdot\mathsf{Im}\left[D\sigma_{\lambda}\left(\mathbf{x}\right)\right],
\]
which is precisely the first part of \eqref{bf-1}. As a consequence,
the functions $F_{i}$'s define a set of $l$ independent first integrals
for $H$. Now, let us prove that they are isotropic. To do that, we
just need the Lemma \ref{lema1}. In fact, combining \eqref{inc}
and \eqref{kerf} we have exactly the inclusion \eqref{sinc}. Finally,
according to Lemma \ref{lema1}, the functions $F_{1},...,F_{l}$
are isotropic. So, given a complete solution, we have constructed
a set of local independent isotropic first integrals around every
point of the phase space. Moreover, the ``inverse construction''
can also be made, as we show below.
\begin{thm}
\label{fcsi}Given a complete solution $\Sigma:\mathbb{R}^{r}\times\mathbb{R}^{l}\rightarrow\mathbb{R}^{2n}$
and a point of $\mathbb{R}^{2n}$, a set of $l$ local independent
and isotropic first integrals is defined by the formula {[}see \eqref{p}{]}
\[
F=\left(F_{1},...,F_{l}\right)\coloneqq\mathfrak{p}\circ\Sigma^{-1},
\]
being $\Sigma^{-1}$ a local inverse of $\Sigma$ around the given
point. Reciprocally, given a set of $l$ independent and isotropic
first integrals $F_{1},...,F_{l}$ and a point of $\mathbb{R}^{2n}$,
a generalized local complete solution with image around such a point
can be constructed.
\end{thm}
\textit{Proof}. The first implication have been proved above. Let
us show the second one. Fix a point $\left(\mathbf{q}^{0},\mathbf{p}^{0}\right)$
of $\mathbb{R}^{2n}$. The independence of the functions $F_{i}$'s
ensures that the $l\times2n$ matrix $DF\left(\mathbf{q}^{0},\mathbf{p}^{0}\right)$
has $l$ columns linearly independent. For simplicity, let us write
$q_{i}=y_{i}$ and $p_{i}=y_{n+i}$. Suppose that the mentioned columns
are
\[
\left(\begin{array}{c}
\frac{\partial F_{1}}{\partial y_{i_{1}}}\\
\frac{\partial F_{2}}{\partial y_{i_{1}}}\\
\vdots\\
\frac{\partial F_{l}}{\partial y_{i_{1}}}
\end{array}\right),\left(\begin{array}{c}
\frac{\partial F_{1}}{\partial y_{i_{2}}}\\
\frac{\partial F_{2}}{\partial y_{i_{2}}}\\
\vdots\\
\frac{\partial F_{l}}{\partial y_{i_{2}}}
\end{array}\right),...,\left(\begin{array}{c}
\frac{\partial F_{1}}{\partial y_{i_{l}}}\\
\frac{\partial F_{2}}{\partial y_{i_{l}}}\\
\vdots\\
\frac{\partial F_{l}}{\partial y_{i_{l}}}
\end{array}\right).
\]
Now, call $z_{r+k}=y_{i_{k}}$, $k=1,...,l$, and call $z_{1},...,z_{r}$
to the rest of $y$'s (in some order). Finally, define $\Pi=\left(\Pi_{1},...,\Pi_{r}\right):\mathbb{R}^{2n}\rightarrow\mathbb{R}^{r}$
such that $\Pi_{i}\left(\mathbf{q},\mathbf{p}\right)=z_{i}$, $i=1,...,r$.
It is easy to see that the function $\left(\Pi,F\right):\mathbb{R}^{2n}\rightarrow\mathbb{R}^{r}\times\mathbb{R}^{l}$
is locally invertible around $\left(\mathbf{q}^{0},\mathbf{p}^{0}\right)$.
It is enough to check that the differential $D\left(\Pi,F\right)\left(\mathbf{q}^{0},\mathbf{p}^{0}\right)$
is a full rank $2n\times2n$ matrix. Let us prove that any local inverse
$\Sigma\coloneqq\left(\Pi,F\right)^{-1}$ is a complete solution.
To simplify the notation, we shall assume that $\Sigma$ is globally
defined. Note first that each function 
\[
\sigma_{\lambda}:\mathbb{R}^{r}\rightarrow\mathbb{R}^{2n}:\mathbf{x}\mapsto\Sigma\left(\mathbf{x},\lambda\right)
\]
 is left inverted by $\Pi$. In fact, 
\[
\left(\Pi,F\right)\left(\sigma_{\lambda}\left(\mathbf{x}\right)\right)=\left(\mathbf{x},\lambda\right),
\]
so $\Pi\circ\sigma_{\lambda}\left(\mathbf{x}\right)=\mathbf{x}$.
It rests to show that each $\sigma_{\lambda}$ satisfies \eqref{bf-1}.
To do that, it is enough to imitate the steps we made above. Since
the functions $F_{i}$'s are first integrals, we know that Eqs. \eqref{fi}
and \eqref{fi-1} hold. From the very definition of $\Sigma$, it
is clear that $F=\mathfrak{p}\circ\Sigma^{-1}$, and consequently
Eq. \eqref{kerf} also holds. Combining \eqref{fi-1} and \eqref{kerf},
we have the first part of \eqref{bf-1} for each $\sigma_{\lambda}$.
Finally, let us prove the second part. The isotropy of $F$ says that
\eqref{sinc} holds, which combined with \eqref{kerf} gives rise
to \eqref{inc}. But the last equation says precisely that the second
part of \eqref{bf-1} is fulfilled, as we wanted to show$.\;\;\;\diamondsuit$

\bigskip{}

\textbf{Example 2.} Consider again the isotropic harmonic oscillator
with $3$ degrees of freedom (see Example 1). As we already know,
the functions $H,H_{1},P_{12},P_{13}$ are independent first integrals.
Moreover, it can be shown that they are isotropic (and do not satisfy
the closure condition) and that $\Pi:\mathbb{R}^{6}\rightarrow\mathbb{R}^{2}$,
given by $\Pi\left(\mathbf{q},\mathbf{p}\right)=\left(q_{1},p_{3}\right)$,
is transverse to $F=\left(H,H_{1},P_{12},P_{13}\right)$, i.e. $\left(\Pi,F\right)$
is a local diffeomorphism. Then, according to the last theorem, every
local inverse defines a generalized complete solution. One of such
inverses $\Sigma=\left(\Sigma^{1},\Sigma^{2}\right):U\subseteq\mathbb{R}^{2}\times\mathbb{R}^{4}\rightarrow V\subseteq\mathbb{R}^{3}\times\mathbb{R}^{3}$
is given by
\begin{equation}
\Sigma^{1}\left(\mathbf{x},\lambda\right)=\left(x_{1},f\left(\mathbf{x},\lambda\right),\frac{x_{1}x_{2}-\lambda_{4}}{\sqrt{\lambda_{2}-x_{1}^{2}}}\right)\label{S1}
\end{equation}
and
\begin{equation}
\Sigma^{2}\left(\mathbf{x},\lambda\right)=\left(\sqrt{\lambda_{2}-x_{1}^{2}},\frac{\lambda_{3}+f\left(\mathbf{x},\lambda\right)\,\sqrt{\lambda_{2}-x_{1}^{2}}}{x_{1}},x_{2}\right),\label{S2}
\end{equation}
where $U$ and $V$ are appropriate open subsets, and $f=f\left(\mathbf{x},\lambda\right)$
is a solution of the quadratic equation
\[
f^{2}+\left(\frac{\lambda_{3}+f\,\sqrt{\lambda_{2}-x_{1}^{2}}}{x_{1}}\right)^{2}+\frac{\left(x_{1}x_{2}-\lambda_{4}\right)^{2}}{\lambda_{2}-x_{1}^{2}}+x_{2}+\lambda_{2}-\lambda_{1}=0.\;\;\;\diamondsuit
\]

\bigskip{}

Concluding, at a local level, having a generalized complete solution
is the same as having a set of isotropic first integrals. As a consequence,
using the results of Section \ref{isotonci}, from a generalized complete
solution we can construct a set of local first integrals that make
our Hamiltonian system, along an open dense subset of the phase space,
a locally NCI system. This gives rise to an extension of the Theorem
\ref{shjt} (and its ``converse,'' commented below it) to the present
generalized context. Also, this tells us that the knowledge of a generalized
complete solution ensures exact solvability. Note however that, according
to the procedures presented so far, in order to construct the trajectories
of the system, such a generalized complete solution is not enough
(we also need the rest of first integrals constructed in Proposition
\ref{fitn}). 

In the next section we shall develop an alternative procedure which
only uses a generalized (local) complete solution. This is because
such a procedure does not rest on the Lie theorem on integrability
by quadratures. Also, the new procedure enable us to find all the
trajectories of the system, not only those contained in a dense subset
of the phase space.

\section{An alternative procedure for integration}

Let us see that a generalized complete solution defines (up to quadratures)
a transformation of the canonical equations into a set of algebraic
equations.
\begin{prop}
Given a generalized complete solution $\Sigma:\mathbb{R}^{r}\times\mathbb{R}^{l}\rightarrow\mathbb{R}^{2n}$
for $H$, we can construct up to quadratures a function $h:\mathbb{R}^{l}\rightarrow\mathbb{R}$
and a function $W:\mathbb{R}^{r}\times\mathbb{R}^{l}\rightarrow\mathbb{R}$
such that
\begin{equation}
H\circ\Sigma\left(\mathbf{x},\lambda\right)=h\left(\lambda\right),\;\;\;\forall\mathbf{x}\in\mathbb{R}^{r},\label{eq:Hh}
\end{equation}
and
\begin{equation}
\theta\left(\Sigma\left(\mathbf{x},\lambda\right)\right)\cdot\frac{\partial\Sigma}{\partial\mathbf{x}}\left(\mathbf{x},\lambda\right)=\frac{\partial W}{\partial\mathbf{x}}\left(\mathbf{x},\lambda\right),\;\;\;\forall\mathbf{x}\in\mathbb{R}^{r},\label{Ss}
\end{equation}
where $\theta:\mathbb{R}^{2n}\rightarrow\mathbb{R}^{2n}:\left(\mathbf{q},\mathbf{p}\right)\mapsto\left(\mathbf{p},0\right)$.
\end{prop}
\textit{Proof}. Suppose that $\Sigma$ is given by a family of partial
solutions $\sigma_{\lambda}$. Since $\nabla\left(H\circ\sigma_{\lambda}\right)=0$
for each $\lambda$ {[}recall \eqref{gradH}{]}, then $H\circ\sigma_{\lambda}\left(\mathbf{x}\right)$
do not depends on $\mathbf{x}$, but only on $\lambda$. This defines
a function $h:\mathbb{R}^{l}\rightarrow\mathbb{R}$ by the formula
\[
h\left(\lambda\right)\coloneqq H\circ\sigma_{\lambda}\left(\mathbf{x}\right)=H\circ\Sigma\left(\mathbf{x},\lambda\right),
\]
as we claim above {[}see \eqref{eq:Hh}{]}. Now, let us construct
$W$. The condition $\left(D\sigma_{\lambda}\right)^{t}\cdot J\cdot D\sigma_{\lambda}=0$,
if we write $\sigma_{\lambda}=\left(\sigma_{\lambda}^{1},\sigma_{\lambda}^{2}\right)$,
says exactly that
\[
\frac{\partial}{\partial x_{k}}\left[\sum_{j=1}^{n}\left(\sigma_{\lambda}^{2}\right)_{j}\frac{\partial}{\partial x_{i}}\left(\sigma_{\lambda}^{1}\right)_{j}\right]-\frac{\partial}{\partial x_{i}}\left[\sum_{j=1}^{n}\left(\sigma_{\lambda}^{2}\right)_{j}\frac{\partial}{\partial x_{k}}\left(\sigma_{\lambda}^{1}\right)_{j}\right]=0,
\]
what implies that 
\[
\sum_{j=1}^{n}\left(\sigma_{\lambda}^{2}\right)_{j}\frac{\partial}{\partial x_{i}}\left(\sigma_{\lambda}^{1}\right)_{j}=\frac{\partial}{\partial x_{i}}W_{\lambda}
\]
for some function $W_{\lambda}$. As it is well-known, each function
$W_{\lambda}$ can be obtained up to quadratures. In terms of the
latter, we have a function $W$ given by the formula $W\left(\mathbf{x},\lambda\right)\coloneqq W_{\lambda}\left(\mathbf{x}\right)$.
Finally, using the function $\theta$ defined above, it is easy to
show that 
\[
\sum_{j=1}^{n}\left(\sigma_{\lambda}^{2}\right)_{j}\left(\mathbf{x}\right)\frac{\partial}{\partial x_{i}}\left(\sigma_{\lambda}^{1}\right)_{j}\left(\mathbf{x}\right)=\left[\left[\theta\left(\sigma_{\lambda}\left(\mathbf{x}\right)\right)\right]^{t}\cdot\frac{\partial}{\partial\mathbf{x}}\sigma_{\lambda}\left(\mathbf{x}\right)\right]_{i},
\]
from which the Eq. \eqref{Ss} easily follows$.\;\;\;\diamondsuit$

\bigskip{}

\begin{rem*}
If the partial solutions $\sigma_{\lambda}$ are of the form \eqref{form},
then
\[
\theta\left(\Sigma\left(\mathbf{q},\lambda\right)\right)\cdot\frac{\partial\Sigma}{\partial\mathbf{q}}\left(\mathbf{q},\lambda\right)=\left(\begin{array}{cc}
\hat{\sigma}_{\lambda}\left(\mathbf{q}\right) & 0\end{array}\right)\cdot\left(\begin{array}{c}
I_{n}\\
\frac{\partial\hat{\sigma}_{\lambda}}{\partial\mathbf{q}}\left(\mathbf{q}\right)
\end{array}\right)=\hat{\sigma}_{\lambda}\left(\mathbf{q}\right).
\]
Thus, the function $W$ in the Eq. \eqref{Ss} is an extension of
the idea of Hamilton's characteristic function.
\end{rem*}
\bigskip{}

\begin{rem*}
In terms of the projection $\mathfrak{p}$ {[}see Eq. \eqref{p}{]},
the function $h$ can be characterized by the equality
\begin{equation}
H\circ\Sigma=h\circ\mathfrak{p}.\label{Hh}
\end{equation}
\end{rem*}
\bigskip{}

\textbf{Example 3. }Coming back to the isotropic harmonic oscillator
(see Examples 1 and 2), for the local complete solution given by the
Eqs. \eqref{S1} and \eqref{S2}, since $F\circ\Sigma\left(\mathbf{x},\lambda\right)=\left(\lambda_{1},...,\lambda_{4}\right)$,
and consequently $H\circ\Sigma\left(\mathbf{x},\lambda\right)=\lambda_{1}$,
it follows that
\[
h\left(\lambda\right)=\lambda_{1}.
\]
On the other hand, the general expression for $W$ is given by the
formula
\[
W\left(\mathbf{x},\lambda\right)=\int_{x_{0,1}}^{x_{1}}\left(\sum_{i=1}^{3}\Sigma_{i}^{2}\,\frac{\partial\Sigma_{i}^{1}}{\partial x_{1}}\right)\left(t,x_{2},\lambda\right)\,dt+\int_{x_{0,2}}^{x_{2}}\left(\sum_{i=1}^{3}\Sigma_{i}^{2}\,\frac{\partial\Sigma_{i}^{1}}{\partial x_{2}}\right)\left(x_{0,1},t,\lambda\right)\,dt.\;\;\;\diamondsuit
\]

\bigskip{}
Now, let us define $\varphi:\mathbb{R}^{r}\times\mathbb{R}^{l}\rightarrow\mathbb{R}^{l}$
as
\begin{equation}
\varphi\left(\mathbf{x},\lambda\right)\coloneqq\frac{\partial W}{\partial\lambda}\left(\mathbf{x},\lambda\right)-\theta\left(\Sigma\left(\mathbf{x},\lambda\right)\right)\cdot\frac{\partial\Sigma}{\partial\lambda}\left(\mathbf{x},\lambda\right).\label{Fi}
\end{equation}

\begin{prop}
\label{finj}The linear map $\frac{\partial\varphi}{\partial\mathbf{x}}\left(\mathbf{x},\lambda\right):\mathbb{R}^{r}\rightarrow\mathbb{R}^{l}$
is injective for all $\left(\mathbf{x},\lambda\right)$. 
\end{prop}
\textit{Proof}. It is easy to see that, for every column vector $v\in\mathbb{R}^{r}$,
\begin{equation}
\begin{array}{lll}
\frac{\partial\varphi}{\partial\mathbf{x}}\left(\mathbf{x},\lambda\right)\cdot v & = & \left(\frac{\partial}{\partial\lambda}\left[\theta\left(\Sigma\left(\mathbf{x},\lambda\right)\right)\right]\cdot\frac{\partial\Sigma}{\partial\mathbf{x}}\left(\mathbf{x},\lambda\right)-\frac{\partial}{\partial\mathbf{x}}\left[\theta\left(\Sigma\left(\mathbf{x},\lambda\right)\right)\right]\cdot\frac{\partial\Sigma}{\partial\lambda}\left(\mathbf{x},\lambda\right)\right)\cdot v\\
\\
 & = & D\mathfrak{p}\left(\mathbf{x},\lambda\right)\cdot\left(D\Sigma\left(\mathbf{x},\lambda\right)\right)^{t}\cdot J\cdot D\Sigma\left(\mathbf{x},\lambda\right)\cdot\left(\begin{array}{c}
v\\
0
\end{array}\right),
\end{array}\label{3}
\end{equation}
where we are using that $D\mathfrak{p}\left(\mathbf{x},\lambda\right)=\left(0_{l\times r}\;I_{l}\right)$
{[}see Eq. \eqref{p}{]}. We know from \eqref{iwv} that 
\begin{equation}
\left(D\Sigma\left(\mathbf{x},\lambda\right)\right)^{t}\cdot J\cdot D\Sigma\left(\mathbf{x},\lambda\right)\cdot\left(\begin{array}{c}
v\\
0
\end{array}\right)=\left(\begin{array}{c}
0\\
v'
\end{array}\right)\label{v00v}
\end{equation}
for some column vector $v'\in\mathbb{R}^{l}$, and consequently
\[
\frac{\partial\varphi}{\partial\mathbf{x}}\left(\mathbf{x},\lambda\right)\cdot v=v'.
\]
If $v'=0$, since $D\Sigma\left(\mathbf{x},\lambda\right)$ and $J$
are bijective linear maps, then $v$ must be zero, what ends our proof$.\;\;\;\diamondsuit$
\begin{prop}
A curve $\left(\mathbf{q}\left(t\right),\mathbf{p}\left(t\right)\right)$
is a trajectory for $H$ with initial condition $\left(\mathbf{q}_{0},\mathbf{p}_{0}\right)$
if and only if, given a local inverse of $\Sigma$ around $\left(\mathbf{q}_{0},\mathbf{p}_{0}\right)$,
the curve $\left(\mathbf{x}\left(t\right),\lambda\left(t\right)\right)\coloneqq\Sigma^{-1}\left(\mathbf{q}\left(t\right),\mathbf{p}\left(t\right)\right)$
satisfies
\begin{equation}
\varphi\left(\mathbf{x}\left(t\right),\lambda_{0}\right)=\varphi\left(\mathbf{x}_{0},\lambda_{0}\right)-t\,Dh\left(\lambda_{0}\right)\;\;\;\textrm{and}\;\;\;\lambda\left(t\right)=\lambda_{0},\label{alg}
\end{equation}
with $\left(\mathbf{x}_{0},\lambda_{0}\right)\coloneqq\Sigma^{-1}\left(\mathbf{q}_{0},\mathbf{p}_{0}\right)$
.
\end{prop}
\textit{Proof}. Consider a trajectory $\left(\mathbf{q}\left(t\right),\mathbf{p}\left(t\right)\right)$
with initial condition $\left(\mathbf{q}_{0},\mathbf{p}_{0}\right)$
and a local inverse $\Sigma^{-1}$ around $\left(\mathbf{q}_{0},\mathbf{p}_{0}\right)$,
and define $\left(\mathbf{x}\left(t\right),\lambda\left(t\right)\right)\coloneqq\Sigma^{-1}\left(\mathbf{q}\left(t\right),\mathbf{p}\left(t\right)\right)$
and $\left(\mathbf{x}_{0},\lambda_{0}\right)\coloneqq\Sigma^{-1}\left(\mathbf{q}_{0},\mathbf{p}_{0}\right)$.
(Of course, $\mathbf{x}\left(0\right)=\mathbf{x}_{0}$ and $\lambda\left(0\right)=\lambda_{0}$).
We known from the previous section that
\[
\lambda\left(t\right)=\mathfrak{p}\circ\Sigma^{-1}\left(\mathbf{q}\left(t\right),\mathbf{p}\left(t\right)\right)
\]
is constant. Then, $\lambda\left(t\right)=\lambda_{0}$. So, it is
enough to see that

\begin{equation}
\frac{d}{dt}\left[\varphi\left(\mathbf{x}\left(t\right),\lambda_{0}\right)\right]^{t}=-\nabla h\left(\lambda_{0}\right).\label{dfi}
\end{equation}
On the one hand, Eq. \eqref{3} tells us that
\begin{equation}
\frac{d}{dt}\left[\varphi\left(\mathbf{x}\left(t\right),\lambda_{0}\right)\right]^{t}=D\mathfrak{p}\left(\mathbf{x}\left(t\right),\lambda_{0}\right)\cdot\left(D\Sigma\left(\mathbf{x}\left(t\right),\lambda_{0}\right)\right)^{t}\cdot J\cdot D\Sigma\left(\mathbf{x}\left(t\right),\lambda_{0}\right)\cdot\left(\begin{array}{c}
v\\
0
\end{array}\right),\label{dfidt}
\end{equation}
with $v^{t}=\dot{\mathbf{x}}\left(t\right)$. On the other hand, using
\eqref{ceq},
\begin{equation}
\begin{array}{lll}
\left(\begin{array}{c}
v\\
0
\end{array}\right) & = & \frac{d}{dt}\left(\mathbf{x}\left(t\right),\lambda\left(t\right)\right)^{t}=D\Sigma^{-1}\left(\mathbf{q}\left(t\right),\mathbf{p}\left(t\right)\right)\cdot\left(\begin{array}{c}
\left(\dot{\mathbf{q}}\left(t\right)\right)^{t}\\
\left(\dot{\mathbf{p}}\left(t\right)\right)^{t}
\end{array}\right)\\
\\
 & = & D\Sigma^{-1}\left(\mathbf{q}\left(t\right),\mathbf{p}\left(t\right)\right)\cdot\left(\begin{array}{c}
\left(\frac{\partial H}{\partial\mathbf{p}}\left(\mathbf{q}\left(t\right),\mathbf{p}\left(t\right)\right)\right)^{t}\\
\left(-\frac{\partial H}{\partial\mathbf{q}}\left(\mathbf{q}\left(t\right),\mathbf{p}\left(t\right)\right)\right)^{t}
\end{array}\right)\\
\\
 & = & \left(D\Sigma\left(\mathbf{x}\left(t\right),\lambda_{0}\right)\right)^{-1}\cdot J\cdot\left(\nabla H\left(\Sigma\left(\mathbf{x}\left(t\right),\lambda_{0}\right)\right)\right).
\end{array}\label{v0}
\end{equation}
Thus, 
\begin{equation}
\begin{array}{l}
D\mathfrak{p}\left(\mathbf{x}\left(t\right),\lambda_{0}\right)\cdot\left(D\Sigma\left(\mathbf{x}\left(t\right),\lambda_{0}\right)\right)^{t}\cdot J\cdot D\Sigma\left(\mathbf{x}\left(t\right),\lambda_{0}\right)\cdot\left(\begin{array}{c}
v\\
0
\end{array}\right)\\
\\
=-D\mathfrak{p}\left(\mathbf{x}\left(t\right),\lambda_{0}\right)\cdot\left(D\Sigma\left(\mathbf{x}\left(t\right),\lambda_{0}\right)\right)^{t}\cdot\nabla H\left(\Sigma\left(\mathbf{x}\left(t\right),\lambda_{0}\right)\right)\\
\\
=-D\mathfrak{p}\left(\mathbf{x}\left(t\right),\lambda_{0}\right)\cdot\nabla\left(H\circ\Sigma\right)\left(\mathbf{x}\left(t\right),\lambda_{0}\right)=-\nabla h\left(\lambda_{0}\right),
\end{array}\label{lp}
\end{equation}
where we have used that 
\begin{equation}
\nabla\left(H\circ\Sigma\right)\left(\mathbf{x},\lambda\right)=\left(D\mathfrak{p}\left(\mathbf{x},\lambda\right)\right)^{t}\cdot\nabla h\left(\lambda\right)\label{hph}
\end{equation}
{[}see Eq. \eqref{Hh}{]} and
\[
D\mathfrak{p}\left(\mathbf{x},\lambda\right)\cdot\left(D\mathfrak{p}\left(\mathbf{x},\lambda\right)\right)^{t}=\left(0_{l\times r}\;I_{l}\right)\cdot\left(\begin{array}{c}
0_{r\times l}\\
I_{l}
\end{array}\right)=I_{l}.
\]
This ends the proof of the first affirmation. For the converse, we
must show the equality {[}see the first row of Eq. \eqref{v0}{]}
\begin{equation}
J\cdot D\Sigma\left(\mathbf{x}\left(t\right),\lambda_{0}\right)\cdot\left(\begin{array}{c}
v\\
0
\end{array}\right)=-\nabla H\left(\Sigma\left(\mathbf{x}\left(t\right),\lambda_{0}\right)\right).\label{fs}
\end{equation}
Combining \eqref{dfi} and \eqref{dfidt} we have that
\[
D\mathfrak{p}\left(\mathbf{x}\left(t\right),\lambda_{0}\right)\cdot\left(D\Sigma\left(\mathbf{x}\left(t\right),\lambda_{0}\right)\right)^{t}\cdot J\cdot D\Sigma\left(\mathbf{x}\left(t\right),\lambda_{0}\right)\cdot\left(\begin{array}{c}
v\\
0
\end{array}\right)=-\nabla h\left(\lambda_{0}\right),
\]
and multiplying by $\left(D\mathfrak{p}\left(\mathbf{x}\left(t\right),\lambda_{0}\right)\right)^{t}$
{[}see Eq. \eqref{hph}{]}
\[
\begin{array}{ll}
\left(D\mathfrak{p}\left(\mathbf{x}\left(t\right),\lambda_{0}\right)\right)^{t}\cdot D\mathfrak{p}\left(\mathbf{x}\left(t\right),\lambda_{0}\right)\cdot\left(D\Sigma\left(\mathbf{x}\left(t\right),\lambda_{0}\right)\right)^{t}\cdot J\cdot D\Sigma\left(\mathbf{x}\left(t\right),\lambda_{0}\right)\cdot\left(\begin{array}{c}
v\\
0
\end{array}\right) & =\\
\\
=-\nabla\left(H\circ\Sigma\right)\left(\mathbf{x}\left(t\right),\lambda_{0}\right)=-\left(D\Sigma\left(\mathbf{x}\left(t\right),\lambda_{0}\right)\right)^{t}\cdot\nabla H\left(\Sigma\left(\mathbf{x}\left(t\right),\lambda_{0}\right)\right).
\end{array}
\]
But, from Eq. \eqref{v00v} and the fact that 
\[
\left(D\mathfrak{p}\left(\mathbf{x},\lambda\right)\right)^{t}\cdot D\mathfrak{p}\left(\mathbf{x},\lambda\right)\cdot\left(\begin{array}{c}
0\\
v'
\end{array}\right)=\left(\begin{array}{c}
0_{r\times l}\\
I_{l}
\end{array}\right)\cdot\left(0_{l\times r}\;I_{l}\right)\cdot\left(\begin{array}{c}
0\\
v'
\end{array}\right)=\left(\begin{array}{c}
0\\
v'
\end{array}\right)
\]
for all column vectors $v'\in\mathbb{R}^{l}$, the equality
\[
\left(D\Sigma\left(\mathbf{x}\left(t\right),\lambda_{0}\right)\right)^{t}\cdot J\cdot D\Sigma\left(\mathbf{x}\left(t\right),\lambda_{0}\right)\cdot\left(\begin{array}{c}
v\\
0
\end{array}\right)=-\left(D\Sigma\left(\mathbf{x}\left(t\right),\lambda_{0}\right)\right)^{t}\cdot\nabla H\left(\Sigma\left(\mathbf{x}\left(t\right),\lambda_{0}\right)\right)
\]
follows, which clearly implies Eq. \eqref{fs}$.\;\;\;\diamondsuit$

\bigskip{}

Summing up, giving a generalized complete solution, we can transform
the canonical Hamilton equations into a set of algebraic equations
{[}see \eqref{alg}{]} whose data can be obtained up to quadratures.
Also, the solutions of the differential and the algebraic equations
are in bijection (according to the last proposition). And finally,
the Proposition \ref{finj} combined with the Implicit Function Theorem
ensures that Eq. \eqref{alg} can be solved for $\mathbf{x}\left(t\right)$,
i.e. the solutions of the algebraic equations can be constructed. 
\begin{thm}
\label{gcses}Given a Hamiltonian system, if we know\footnote{As we emphasized in Remark \ref{cjl}, we are asking the knowledge
of a complete solution, not just its existence. } a generalized complete solution solution for it, then such a system
is exactly solvable (along the whole of the phase space).
\end{thm}
Thus, given a generalized complete solution, the trajectories of the
system can be constructed up to quadratures by following a procedure
different to that related to the NCI, which can be described as follows.
Given $\Sigma$:
\begin{enumerate}
\item construct (up to quadratures) the functions $h$ and $W$ from Eqs.
\eqref{eq:Hh} and \eqref{Ss}, respectively;
\item construct from Eq. \eqref{Fi} the function $\varphi$;
\item fix a pair $\left(\mathbf{x}_{0},\lambda_{0}\right)$ and solve the
Eq. \eqref{alg} for $\mathbf{x}\left(t\right)$ by inverting $\varphi$;
\item define $\left(\mathbf{q}\left(t\right),\mathbf{p}\left(t\right)\right)=\Sigma\left(\mathbf{x}\left(t\right),\lambda_{0}\right)$.
\end{enumerate}
Theorem above combined with Theorem \ref{fcsi} constitutes our second
proof of the main assertion of this paper: ``for a given Hamiltonian
system, the knowledge of a set of isotropic first integrals is enough
to integrate its equations of motion up to quadratures.'' Unlike
our first proof, the closure condition do not appear in any step of
the present one.

In parallel with Remark \ref{nonex}, we can see Theorem \ref{gcses}
as another criterium for exact solvability, weaker than NCI. It is
also true in this case (because of Proposition \ref{fitn} and Theorem
\ref{fcsi}) that such a criterium does not give rise to new exactly
solvable systems (different from the NCI ones). Nevertheless, the
theoretical contribution of the theorem is undeniable. We think that
its practical contribution lies on its proof, which constitutes an
integration procedure completely different from the usual ones (where,
in particular, Lie's theorem is not involved at all).

\section*{Conclusions}

In order to summarize the results of this work, given a Hamiltonian
system defined by $H$, we can say that, if we have a set of independent
isotropic first integrals for $H$, we can construct its trajectories
by following two ways:

\smallskip{}

\textbf{based on the Lie integrability theorem}:
\begin{itemize}
\item following the steps of the proof of Proposition \ref{fitn}, construct
the subset $D$ and, around each point of $D$, construct the functions
that makes the system locally NCI;
\item using such functions, apply the points 1, 2 and 3 described at the
end of Section \ref{fnci}.
\end{itemize}
\smallskip{}

\textbf{based on the generalized Hamilton-Jacobi theory}:
\begin{itemize}
\item following the steps of the proof of Theorem \ref{fcsi}, construct
around each point of the phase space a generalized local complete
solution of the HJE;
\item using such solutions, apply the points 1, 2, 3 and 4 described below
Theorem \ref{gcses}.
\end{itemize}

\section*{Appendix}

The Carathéodory-Jacobi-Lie theorem (see for instance \cite{lm})
says that, given independent functions $f_{1},...,f_{r}:\mathbb{R}^{2n}\rightarrow\mathbb{R}$
in involution and a point $\mathbf{x}_{0}\in\mathbb{R}^{2n}$, there
exist an open neighborhood $A$ of $\mathbf{x}_{0}$ and another functions
$f_{r+1},...,f_{n},g_{1},...,g_{n}:A\rightarrow\mathbb{R}$ such that
the whole set of functions (the original together the new ones) are
independent and satisfy
\[
\left\{ f_{i},f_{j}\right\} =\left\{ g_{i},g_{j}\right\} =0\;\;\;\textrm{and}\;\;\;\left\{ f_{i},g_{j}\right\} =\delta_{ij},
\]
i.e. the functions $f_{1},...,f_{n},g_{1},...,g_{n}$ give local canonical
coordinates around $\mathbf{x}_{0}$. Suppose that $r=1$. Then, for
every $l$ such that $n\leq l\leq2n$, the $l$ functions 
\[
F_{1}\coloneqq f_{1},...,F_{n}\coloneqq f_{n},F_{n+1}\coloneqq g_{2n-l+1},...,F_{l}\coloneqq g_{n}
\]
fulfill the conditions $\left(1\right)$ to $\left(4\right)$ of NCI
for the Hamiltonian function $f_{1}$. Condition $\left(1\right)$
and $\left(2\right)$ are immediate. Let us show $\left(3\right)$
and $\left(4\right)$. The brackets $\left\{ F_{i},F_{j}\right\} $
define the $l\times l$ constant matrix
\[
B\coloneqq\left(\begin{array}{ccc}
0_{\left(2n-l\right)\times\left(2n-l\right)} & 0_{\left(2n-l\right)\times\left(l-n\right)} & 0_{\left(2n-l\right)\times\left(l-n\right)}\\
0_{\left(l-n\right)\times\left(2n-l\right)} & 0_{\left(l-n\right)\times\left(l-n\right)} & I_{\left(l-n\right)\times\left(l-n\right)}\\
0_{\left(l-n\right)\times\left(2n-l\right)} & -I_{\left(l-n\right)\times\left(l-n\right)} & 0_{\left(l-n\right)\times\left(l-n\right)}
\end{array}\right),
\]
where $0_{a\times b}$ is the $a\times b$ null matrix and $I_{a\times a}$
is the $a\times a$ identity matrix. The dimension of the kernel of
such a matrix is precisely $2n-l$, hence the condition $\left(3\right)$
is satisfied. Also, defining the functions $P_{ij}:\mathbb{R}^{l}\rightarrow\mathbb{R}$
as $P_{ij}\left(c_{1},...,c_{l}\right)=B_{ij}$, with $1\leq i,j\leq l$,
we have that 
\[
P_{ij}\left(F_{1}\left(\mathbf{x}\right),...,F_{l}\left(\mathbf{x}\right)\right)=B_{ij}=\left\{ F_{i},F_{j}\right\} \left(\mathbf{x}\right),\;\;\;\forall\mathbf{x}\in\mathbb{R}^{2n},
\]
from which the condition $\left(4\right)$ holds. 

\smallskip{}

Summarizing, as a direct consequence of the Carathéodory-Jacobi-Lie
theorem we have the following result. 
\begin{thm*}
Given a function $H:\mathbb{R}^{2n}\rightarrow\mathbb{R}$ and a point
$\mathbf{x}_{0}\in\mathbb{R}^{2n}$, for every $l$ such that $n\leq l\leq2n$,
there exist functions $F_{1},...,F_{l}$ satisfying the NCI conditions
for $H$.
\end{thm*}
Note that above theorem does not meant that every Hamiltonian system
is locally NCI, because about the functions $F_{1},...,F_{l}$ we
simply know that they exist, but we do not have, in principle, any
concrete expression of them. That is to say, we \textit{do not know}
them in principle. What the theorem above says is that, in some sense,
every Hamiltonian systems is ``potentially'' NCI around every point
of its phase space. 

\smallskip{}

Above result is rather known, but we choose to include it in the paper
because we do not know any reference where its proof explicitly appears.

\section*{Acknowledgements}

The author thanks CONICET for its financial support.

\end{document}